\documentclass[sigconf]{acmart}

\usepackage{algorithm}
\usepackage{algorithmic}
\usepackage{amsmath}
\usepackage{siunitx}
\usepackage{graphicx}
\usepackage{caption} 
\usepackage{subcaption}

\AtBeginDocument{%
  }

\copyrightyear{2025}
\acmYear{2025}
\setcopyright{acmlicensed}
\acmConference[MM '25] {Proceedings of the 33rd ACM International Conference on Multimedia}{October 27--31, 2025}{Dublin, Ireland.}
\acmBooktitle{Proceedings of the 33rd ACM International Conference on Multimedia (MM '25), October 27--31, 2025, Dublin, Ireland}
\acmISBN{979-8-4007-2035-2/2025/10}
\acmDOI{10.1145/3746027.3754814}

\begin{document}

\title{FedBAP: Backdoor Defense via Benign Adversarial Perturbation in Federated Learning}


\author{Xinhai Yan}
\affiliation{%
  \institution{School of Cyber Science and Engineering, Wuhan University}
  \city{Wuhan}
  \country{China}}
\email{yanxinhai@whu.edu.cn}

\author{Libing Wu}
\authornote{Corresponding authors.}
\affiliation{%
  \institution{School of Cyber Science and Engineering, Wuhan University}
  \city{Wuhan}
  \country{China}}
\email{wu@whu.edu.cn}

\author{Zhuangzhuang Zhang}
\authornotemark[1]
\affiliation{%
  \institution{School of Cyber Science and Engineering, Wuhan University}
  \city{Wuhan}
  \country{China}}
\email{zhzhuangzhuang@whu.edu.cn}

\author{Bingyi Liu}
\affiliation{%
  \institution{Department of Computer and Artificial Intelligence, Wuhan University of Technology}
  \city{Wuhan}
  \country{China}}
\email{byliu@whut.edu.cn}

\author{Lijuan Huo}
\affiliation{%
  \institution{School of Cyber Science and Engineering, Wuhan University}
  \city{Wuhan}
  \country{China}}
\email{lijuanhuo@whu.edu.cn}

\author{Jing Wang}
\affiliation{%
  \institution{School of Software Engineering, Huazhong University of Science and Technology}
  \city{Wuhan}
  \country{China}}
\email{cswjing@hust.edu.cn}

\renewcommand{\shortauthors}{Xinhai Yan et al.}

\begin{abstract}
  Federated Learning (FL) enables collaborative model training while preserving data privacy, but it is highly vulnerable to backdoor attacks. Most existing defense methods in FL have limited effectiveness due to their neglect of the model's over-reliance on backdoor triggers, particularly as the proportion of malicious clients increases. In this paper, we propose FedBAP, a novel defense framework for mitigating backdoor attacks in FL by reducing the model’s reliance on backdoor triggers. Specifically, first, we propose a perturbed trigger generation mechanism that creates perturbation triggers precisely matching backdoor triggers in location and size, ensuring strong influence on model outputs. Second, we utilize these perturbation triggers to generate benign adversarial perturbations that disrupt the model's dependence on backdoor triggers while forcing it to learn more robust decision boundaries. Finally, we design an adaptive scaling mechanism to dynamically adjust perturbation intensity, effectively balancing defense strength and model performance. The experimental results demonstrate that FedBAP reduces the attack success rates by 0.22\%-5.34\%, 0.48\%-6.34\%, and 97.22\%-97.6\% under three types of backdoor attacks, respectively. In particular, FedBAP demonstrates outstanding performance against novel backdoor attacks.
\end{abstract}



\maketitle

\section{Introduction}

As data privacy and security concerns become increasingly prominent, Federated Learning (FL) ~\cite{mcmahan2017communication} has emerged as a promising distributed machine learning paradigm, attracting widespread attention. Unlike traditional centralized training, FL enables multiple clients to train models locally and upload only model updates to a central server for aggregation, thereby facilitating collaborative learning across devices or institutions while preserving data privacy ~\cite{10870877}. Since FL allows efficient training without exposing raw data, it has been widely adopted in various domains, including smart devices ~\cite{hard2018federated}, healthcare ~\cite{sheller2018multi}, IoT device ~\cite{nguyen2019diot}, and financial risk management ~\cite{cheng2021secureboost}. 

However, this decentralized training paradigm also introduces new security challenges, particularly backdoor attacks ~\cite{gu2017badnets}, which pose a significant threat to FL systems. 
Specifically, during the FL training process, malicious clients can inject stealthy backdoors into local models through methods such as data poisoning or model replacement. These backdoors can then propagate to the global model through the server’s model aggregation. Once the attack succeeds, the global model will output the attacker’s designated target class when encountering a specific trigger, while maintaining high accuracy on normal inputs, making the attack extremely difficult to detect and defend against~\cite{tolpegin2020data,sun2019can,bagdasaryan2020backdoor}. Furthermore, due to the decentralized nature of FL, the server is unable to directly access or audit the clients’ training processes, making backdoor attacks even more elusive and harder to detect \cite{gu2017badnets}.

Existing defense mechanisms can be broadly categorized into two approaches  ~\cite{qin2024resisting}: mitigating the impact of infected models ~\cite{zhang2022flip,zhu2023leadfl,jia2023fedgame,huang2023lockdown,rieger2022crowdguard,nguyen2022flame} and filtering out malicious models or parameters ~\cite{cao2020fltrust,wan2023four,gong2023agramplifier,zhang2022fldetector,wang2022flare,kabir2024flshield, qin2024resisting, huang2024parameter, zhang2022lsfl, yang2024roseagg, li2024backdoorindicator}. However, both of these approaches have significant limitations. The former is challenged by the difficulty of precisely identifying backdoor features, often relying on intrusive interventions that can disrupt the model's ability to learn normal tasks, leading to a decline in main task performance. The latter typically relies on feature statistics or anomaly detection, making it susceptible to failures when dealing with complex or stealthy backdoor strategies, potentially mislabeling legitimate updates, and impacting system stability. Furthermore, due to the non-independent and identically distributed (non-IID) nature of client data in FL, these defense strategies struggle to accommodate the diverse backdoor attack patterns encountered by different clients. Therefore, existing backdoor defense methods still face the following challenges:

\textbf{Challenge 1 (Identification of Backdoor Attack Features)}. The stealthy characteristics of backdoor features fundamentally limit current defenses' capability in both accurate feature identification and subsequent complete backdoor removal.

\textbf{Challenge 2 (Balancing Backdoor Attack Defense and Main Task Accuracy)}. Current backdoor defense methods require excessive interference with the model aggregation process for backdoor removal, consequently degrading the main task's accuracy.

To address the above challenge, we focus on the nature of backdoor attacks, namely that backdoor attacks cause the model to overly rely on triggers rather than the global semantic features. Driven by this insight, we propose FedBAP, a Benign Adversarial Perturbation based defense framework against backdoor attacks in FL. Our goal is to eliminate the model's dependence on backdoor triggers, thereby bypassing the challenging process of backdoor identification and ensuring the model's exclusive focus on globally representative features. To achieve this, we first propose a perturbation trigger generation mechanism that significantly influences the model's backdoor decision boundary by generating perturbation triggers that are very similar in position, and size to the backdoor triggers. Second, we design benign adversarial perturbations generation mechanism that utilize perturbation triggers for adversarial training on the client, to reduce the model's dependence on backdoor triggers and encourage it to learn more robust data features, thereby addressing \textbf{Challenge 1}. Finally, we propose an adaptive scaling mechanism that achieves a balance between defense strength and model performance by dynamically adjusting the intensity of benign adversarial perturbations, thus addressing \textbf{Challenge 2}.
Experimental results show that FedBAP effectively mitigates backdoor attacks while preserving model accuracy, demonstrating its effectiveness as a defense strategy.

Our main contributions are as follows:
\begin{itemize}
    \item We propose FedBAP, a Benign Adversarial Perturbation based defense framework against backdoor attacks in FL. FedBAP effectively mitigates backdoor threats while preserving the model's main task performance.

    \item We propose a novel perspective that introduces a fundamentally new defense approach rooted in the model's training behavior itself, focusing on the essential dependency nature of models on backdoor features, thereby effectively eliminating backdoors.

    \item We conduct extensive experiments under various federated backdoor attack scenarios, demonstrating that FedBAP significantly outperforms existing defense methods and, in some cases, even enhances the model’s accuracy on the main task.
\end{itemize}

\section{Releted Work}

Extensive research ~\cite{lyu2024lurking, nguyen2023iba, liu2024beyond, bagdasaryan2020backdoor, xie2019dba, lyu2023poisoning, li20233dfed, fang2023vulnerability, zhuang2023backdoor, li2024darkfed, zhang2022neurotoxin, shi2024towards} has demonstrated that FL is highly susceptible to backdoor attacks. To mitigate these threats, researchers have proposed various defense mechanisms. Existing work on defending against targeted attacks in FL can be broadly categorized into two main approaches  ~\cite{qin2024resisting}: mitigating the impact of infected models and filtering out infected models or parameters.

\textbf{Mitigating the Impact of Infected Models.}
Several methods aim to reduce the influence of compromised models without explicitly removing them. FLIP ~\cite{zhang2022flip} produces an adversarial training-based approach to generate reverse-triggered augmented data, mitigating backdoor threats by exposing the model to counterexamples. LeadFL ~\cite{zhu2023leadfl} enhances model resilience through Hessian-based regularization, improving gradient stability and reducing the impact of malicious updates. FedGame ~\cite{jia2023fedgame} formulates a game-theoretic defense that models the interaction between defenders and attackers, optimizing global model updates by reverse-engineering backdoor triggers and target classes. Lockdown ~\cite{huang2023lockdown} isolates malicious parameters using subspace training, which includes initialization, search, pruning, and aggregation to defend against backdoor attacks while reducing computational complexity. CrowdGuard ~\cite{rieger2022crowdguard} further improves backdoor resilience by leveraging client feedback, analyzing hidden-layer neuron behavior, and performing iterative pruning to remove poisoned models. FLAME ~\cite{nguyen2022flame} employs dynamic model clustering, adaptive weight clipping, and differential privacy noise injection to detect and suppress adversarial updates.

\textbf{Filtering Out Infected Models or Parameters.}
Another line of work focuses on identifying and excluding compromised updates through robust aggregation and anomaly detection. FLTrust ~\cite{cao2020fltrust} adopts a trust-based framework by leveraging a small clean dataset to assign trust scores to client updates, ensuring that malicious contributions are effectively downweighted during aggregation. FPD ~\cite{wan2023four} produces a four-module defense mechanism incorporating reliable client selection, similarity-based anomaly detection, and update denoising to filter out both colluding and non-colluding adversarial clients. AGRAMPLIFIER ~\cite{gong2023agramplifier} strengthens Byzantine-robust aggregation rules by employing local update amplification techniques, improving robustness against adversarial updates. FLDetector ~\cite{zhang2022fldetector} utilizes consistency detection, Euclidean distance-based suspicious scoring, and k-means clustering to isolate malicious clients. FLARE ~\cite{wang2022flare} analyzes penultimate layer representations to assess update credibility and mitigate poisoning effects. FLShield ~\cite{kabir2024flshield} addresses verification challenges in FL through representative model generation and class-wise loss impact measurement, ensuring the integrity of updates before aggregation. BackdoorIndicator ~\cite{li2024backdoorindicator} injects OOD-based indicator tasks into the global model to proactively detect any backdoor-poisoned local model uploads.

These approaches have proven effective in various settings. However, they overlook the fundamental characteristic of backdoor attacks, where the model develops a strong dependence on backdoor triggers. As the intensity of backdoor attacks increases, the defensive performance of these methods will significantly decrease.

\section{Preliminaries}

\subsection{Federated Learning}

We consider a standard FL setup where a set of clients $S$ aim to collaboratively train a global model $w$ with the coordination of a server. Let $\mathcal{D}_{i}$ be the private training dataset held by the client $i$, where $i \in S$. In the $t$-th communication round, the server first randomly selects a set of clients $S_{t}$, where $\lvert S_{t} \rvert \leq \lvert S \rvert$. The server then distributes the current version of the global model $w_t$ to the selected clients. Each selected client $i \in S_t$ first uses the global model to initialize its local model, then trains its local model on its local training dataset, and finally uploads the local model update to the server. We use $g_{t}^i$ to denote the local model update of the client $i$ in the $t$-th
communication round. The server aggregates the received updates on model weights and updates the current global model weights as follows:

\begin{equation}
    w_{t+1} = w_{t} + \mathcal{A}(\{g_{t}^i | i \in S_{t} \})
\end{equation}
where $\mathcal{A}$ is an aggregation rule adopted by the server. For instance, a widely used aggregation rule FedAvg \cite{mcmahan2017communication} takes an average over the local model updates uploaded by clients.

\subsection{Threat Model}

\textbf{Attacker’s goals.} As the existing studies on FL backdoor attacks, an attacker’s goal is to enforce models to classify data samples with triggers embedded to specific incorrect labels while keeping a high accuracy for samples without triggers embedded. 

\textbf{Attacker’s capabilities.} We assume that the server is honest. Following threat models in previous studies ~\cite{lyu2023poisoning, zhang2022neurotoxin, fang2023vulnerability, xie2019dba, bagdasaryan2020backdoor, zhang2023a3fl}, we consider an attacker that can compromise a certain number of clients. Specifically, the attacker can access to the training datasets and global model updates of these compromised clients, allowing manipulation of their uploaded updates. Furthermore, malicious clients under the attacker’s control can communicate and synchronize their attack strategies. The attacker also has access to a snapshot of the global model in each round and can directly modify both model weights and datasets on compromised clients.

\section{Method}

\begin{figure}[!htbp]
  \centering
  \includegraphics[width=\linewidth]{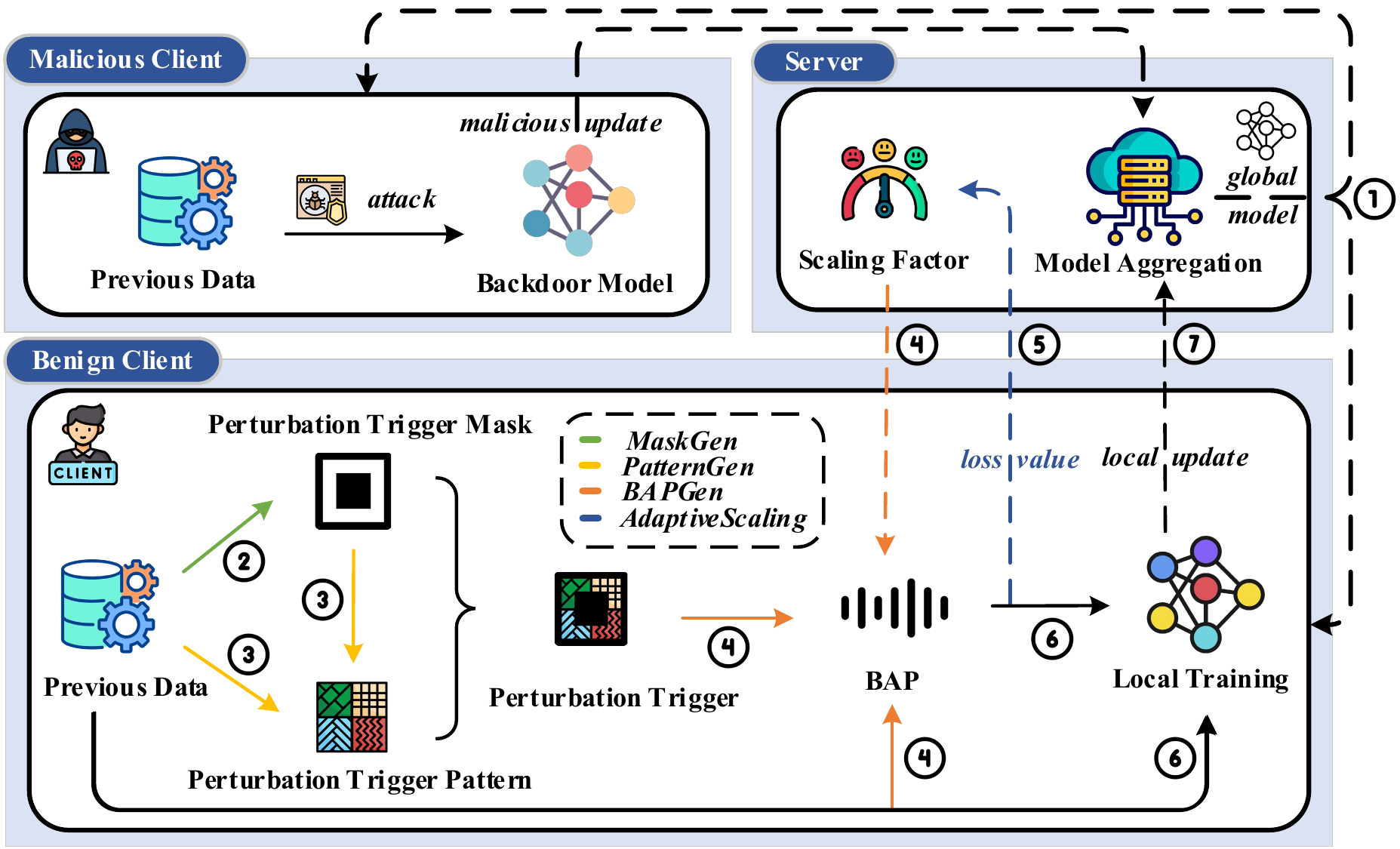}
  \caption{Overview of FedBAP. At the defense start round, the server distributes the global model to clients (\textcircled{1}). Clients generate perturbation triggers (\textcircled{2} and \textcircled{3}), apply BAPGen to create perturbations and upload loss values (\textcircled{4}). The server updates the scaling factor (\textcircled{5}). Clients then train locally and send model updates to the server (\textcircled{6} and \textcircled{7}).}
  \label{FedBAP}
  \Description{}
\end{figure}

\subsection{High-level Description}

As shown in Figure \ref{FedBAP}, FedBAP consists of three key mechanisms: Perturbation Triggers Generation, Benign Adversarial Perturbation Generation, and Adaptive Scaling. The detailed workflow of FedBAP is described in Algorithm \ref{alg:algorithm1}, and the complete process is outlined as follows:

\begin{itemize} 
    \item \textbf{Step 1: Initialization.} Before the start of FL, the server initializes the global model and the scaling factor, which are then distributed to all clients. The scaling factor is a dynamic parameter produced to adjust the strength of benign adversarial perturbations  after the defense is activated.
    
    \item \textbf{Step 2: Perturbation Trigger Generation.} At the designated start round, each client executes the \textit{MaskGen} and \textit{PatternGen} algorithms based on the preliminary converged global model to generate perturbation triggers. These triggers are later used to guide the generation of benign adversarial perturbation.
       
    \item \textbf{Step 3: Benign Adversarial Perturbation Generation.} After the defense starts, clients use the perturbation triggers and the current scaling factor to execute the \textit{BAPGen} algorithm, generating benign adversarial perturbations.  And benign adversarial perturbation loss values are also uploaded once defense is activated.

    \item \textbf{Step 4: Local Training.} Each client trains a local model using its private data and subsequently uploads the corresponding model update. 
    
    \item \textbf{Step 5: Adaptive Scaling.} After the defense is triggered, the server adjusts the scaling factor by executing the \textit{AdaptiveScaling} algorithm based on the received Benign adversarial perturbation loss values and the current scaling factor.
    
    \item \textbf{Step 6: Model Aggregation and Distribution.} The server aggregates local updates into a new global model and distributes it to clients.
\end{itemize}

\begin{algorithm}[tb]
\caption{Overview of FedBAP}
\label{alg:algorithm1}
\raggedright
\textbf{FedBAP:} \\
\hspace*{1em}\textbf{Parameter:} Local epochs $E_1$, learning rate $\eta_1$, start round $t_s$, scaling factor $c_{t_s}$, number of clients $m$, number of communication rounds $N$. \\
\hspace*{1em}\textbf{Output:} Model $w_N$
\begin{algorithmic}[1] 
\STATE init $w_{0}$
\FOR{$t$ = 1 to $N$}
\STATE $S_{t} \gets $ a random set of $m$ clients
\STATE \textit{// Client Side;}
\IF{$t=t_s$}
\STATE $M \gets \textit{MaskGen}(w_t)$
\STATE $\Delta x \gets \textit{PatternGen}(w_t, M)$
\ENDIF
\FOR{$i \in S_{t}$ \textbf{in parallel}}
\STATE $w^\prime \gets w_t$
\STATE $loss_{t}^i \gets 0$
\IF{$t \geq t_s$}
\STATE $loss_{t}^i, w^\prime \gets \textit{BAPGen}(c_{t}, w^\prime, i, M, \Delta x_{i})$
\ENDIF
\FOR{$e$ = 1 to $E_1$}
\FOR{$(x, y) \in \mathcal{D}_i$}
    \STATE $w^\prime \gets w^\prime - \eta_1 \nabla_{w^\prime} \ell((w^\prime(x), y)$
\ENDFOR
\ENDFOR
\STATE $g_{t}^i \gets w^\prime - w_{t}$
\ENDFOR
\STATE \textit{// Server Side;}
\IF{$t \geq t_s$}
\STATE $loss_{t} \gets \frac{1}{\lvert S_{t} \rvert} \sum_{i \in S_{t}}loss_{t}^{i}$
\STATE $c_{t+1} \gets \textit{AdaptiveScaling}(loss_{t}, c_t, t)$
\ENDIF
\STATE $w_{t+1} \gets w_{t} + \frac{1}{\lvert S_{t} \rvert} \sum_{i \in S_{t}}g_{t}^i$
\ENDFOR 
\STATE \textbf{return} $w_N$
\end{algorithmic}
\end{algorithm}

\subsection{Perturbation Trigger Generation}

The main objective of the perturbation trigger generation mechanism is to create a trigger that matches the size and position of the backdoor trigger, but has a greater impact on the model predictions.
Unlike malicious triggers, perturbation triggers are designed to steer gradient updates away from backdoor-reliant behaviors and toward more robust feature representations. According to DEFINITION 1, each perturbation trigger consists of a mask that defines its spatial region and a pattern that specifies the perturbation applied. These components are carefully derived to maintain consistency with attacker-defined triggers while maximizing their ability to disrupt shortcut learning. The following section details the construction methodology.

DEFINITION 1. \textit{The perturbation trigger consists of two components  ~\cite{wang2019neural}:}
\begin{itemize}
    \item \textit{$M$, which represents the trigger mask. It is a two-dimensional matrix with the same height and width as the original image, where the same mask value is applied across all color channels. The values of $M$ range from 0 to 1, determining the proportion of the original image that the trigger can cover.}
    
    \item \textit{$\Delta x$, which represents the trigger pattern. It is a three-dimensional matrix with the same height, width, and number of channels as the original image.}
\end{itemize}

\textit{Let $A(\cdot)$ denote the process of embedding a trigger into the original image $x$. The embedding operation $A(x,M,\Delta x)$ is defined as:}
\begin{equation}
    A(x,M,\Delta x) = (1 - M) \cdot x + M \cdot \Delta x
\end{equation}
\textit{where the mask $M$ selectively mixes the original image $x$ with the perturbation trigger pattern $\Delta x$.}

\textbf{Perturbation Trigger Mask Generation.} 
The first step is to obtain the mask of the perturbation trigger. Ideally, this mask should closely resemble that of the backdoor trigger, ensuring that it accurately interferes with the model’s decision path during adversarial training without straying from the critical region of the backdoor feature. This alignment guarantees that the perturbation effectively aids the model in correcting its over-reliance on the backdoor trigger. According to DEFINITION 2, in a backdoored model, the backdoor trigger has a significantly smaller backdoor distance compared to non-backdoor triggers. Thus, our goal is to identify a trigger that successfully activates the backdoor while minimizing the backdoor distance. Formally, our optimization objective is defined as:

\begin{equation}
\underset{M, \ \Delta x}{\arg\min} \ \ell\left(w(A(x, M, \Delta x)), y_t\right) + \lambda \cdot \| M \| \label{eq:1}
\end{equation}
where $\ell(\cdot)$ denotes the cross-entropy loss function, $w(\cdot)$ represents the neural network, $x$ is the input image, $M$ is the trigger mask, $\Delta x$ is the trigger pattern, $y_t$ is the target class, $\lambda$ is a regularization coefficient controlling the trade-off between classification loss and the sparsity of the mask $M$.

DEFINITION 2. \textit{Given an image from the source class $i$, a trigger composed of a mask $M$ and a pattern $\Delta x$ can flip the image’s label to the target class $j$. The class-wise distance $d_{i \to j}$ is measured as $\| M_{i \to j}\|$, where $\| \cdot \|$ denotes the $L_1$ norm, representing the number of pixels that need to be modified to flip the label from class $i$ to class $j$. This metric quantifies the difficulty of transitioning from class $i$ to class $j$ ~\cite{wang2019neural}. Based on this, we introduce the concept of backdoor distance. Given an image from an arbitrary source class, if a backdoor exists, the backdoor distance $d_{\forall \to t}$ is defined as $\| M_{t}\|$, where class $t$ is the target class. Since the existence of the backdoor makes it easier to flip samples from any class to the target class, we have the inequality:}

\begin{equation}
\| M_{t}\| \ll \| M_{i \to j}\|, \ \text{s.t.} \ j \neq t
\end{equation}
\textit{indicating that flipping a sample to the backdoor target class is significantly easier than flipping it to any other classes.}

Each client iterates over all possible target classes and optimizes the mask $M$ according to Equation \ref{eq:1}. At the end of each epoch, the client evaluates the current trigger mask. If the backdoor accuracy of the current trigger is not lower than the predefined threshold, and the backdoor distance of the current mask is smaller than the client's best recorded mask, then the client's best mask is updated to the current mask. Once all clients have obtained their optimal masks, the server aggregates the masks by computing the average mask across all clients. The server then binarizes the aggregated mask by setting values greater than or equal to 0.5 to 1, and values less than 0.5 to 0. The detailed procedure of the Perturbation Trigger Mask Generation Algorithm \textit{MaskGen} is presented in the appendix.

\textbf{Perturbation Trigger Pattern Generation.} 
After obtaining the mask, the next step is to determine the perturbation trigger pattern. We aim to maximize the perturbation trigger's impact on the model’s output by increasing the difference in the penultimate layer representations (PLR). According to DEFINITION 3, differences in the PLR space translate into corresponding differences in output probabilities. Therefore, when the distance between two PLRs is large, the corresponding output probability distributions exhibit significant divergence. If the perturbation trigger maximally perturbs the PLR, it can induce a substantial change in the model’s output.

DEFINITION 3. \textit{Let the neural network be defined as $f: \mathbb{R}^{d_1 \times d_2 
\times d_3} \\ \to \mathbb{R}^{c}$, mapping an input $x \in \mathbb{R}^{d_1 \times d_2 
\times d_3}$ to a $c$-dimensional probability vector $q \in \mathbb{R}^{c}$, which $c$ is the number of classes. We denote the mapping from the input to the penultimate layer representations (PLR) as $g:\mathbb{R}^{d_1 \times d_2 \times d_3} \to \mathbb{R}^{o}$. The output of $g$ is the PLR, denoted as $r \in \mathbb{R}^{o}$ ~\cite{wang2022flare}. Finally, we define $\sigma: r \in \mathbb{R}^{o} \to q \in \mathbb{R}^{c}$ as the mapping function from the PLR to the output probability vector. We use $\Omega = [\omega_{1}, \omega_{2}, \cdots, \omega_{c}]$ to represent the weight connecting the penultimate layer to the last layer where $\omega_{k} \in \mathbb{R}^{o}$ denotes the weights connecting to the $k$-th neuron of the output layer. We have}

\begin{equation}
    \left\| q^{1}-q^{2} \right\|_{2} \leq \left\| \Omega \right\|_{2} \left\| r_{1}-r_{2} \right\|_{2}
\end{equation}
\textit{where $r_1$ and $r_2$ are the PLR of two input $x_1$ and $x_2$ respectively, $q^{1}$ and $q^{2}$ are the output probability vector for input $x_1$ and $x_2$ respectively, $\left\| q^{1}-q^{2} \right\|_{2}$ denotes the Euclidean distance between $q^{1}$ and $q^{2}$, $\left\| r_{1}-r_{2} \right\|_{2}$ denotes the Euclidean distance between $r_{1}$ and $r_{2}$.}

In FL, client data is typically non-IID, causing backdoor behaviors to vary across clients. As a result, each client must independently generate a customized perturbation trigger using local data. To optimize the trigger's pattern, we leverage the PLR, which reflects the model’s output before the final prediction. Let $r_1$ and $r_2$ be the PLRs before and after applying the trigger. We minimize their cosine similarity to maximize the trigger's impact, encouraging it to induce a strong shift in the model’s output. This drives the model to react to the trigger, facilitating the suppression of backdoor reliance. Once optimized, the pattern is stored for generating benign adversarial perturbations. The full procedure is detailed in the \textit{PatternGen} algorithm in the appendix.

\subsection{Benign Adversarial Perturbation Generation}

To effectively defend against backdoor attacks, we are dedicated to breaking the model's reliance on backdoor triggers and encouraging it to learn the global features of the data. To achieve this goal, we design a benign adversarial perturbation generation mechanism. This mechanism leverages perturbation triggers generated by our trigger generation module to conduct adversarial training on clients, thereby reducing the model's reliance on backdoor triggers and encouraging the learning of more robust data features.

Benign adversarial perturbations differ fundamentally from traditional adversarial perturbations. While the latter are crafted to degrade model performance, the former aim to refine the model’s decision boundary and enhance robustness against backdoor attacks. This is achieved by steering the model away from spurious, backdoor-associated features and promoting the learning of more robust and generalizable representations. By persistently applying pressure against backdoor dependencies, benign perturbations reduce the model’s reliance on shortcut-based decision rules and strengthen its overall resilience.

To generate benign adversarial perturbation in practice, we introduce a client-side training algorithm that incorporates perturbation triggers into the local data. Each client embeds the predefined perturbation trigger into a subset of clean samples and enforces correct label prediction through adversarial training. In this process, the perturbation acts as a targeted optimization signal that reshapes the model's internal representations. To further amplify its effectiveness, we apply a scaling factor to control the perturbation’s magnitude. This factor ensures that the perturbation maintains sufficient influence during training, particularly when the model begins adapting and the adversarial loss starts to decrease. Finally, the algorithm returns both the hardened model and the loss value of the benign adversarial perturbation. The detailed procedure of our Benign Adversarial Perturbation Generation \textit{BAPGen} algorithm is presented in the appendix.

\subsection{Adaptive Scaling}

As training progresses, the impact of benign adversarial perturbation tends to weaken. This is because benign adversarial perturbation is essentially a gradient update generated through adversarial training, and as the model converges, the corresponding adversarial loss value decreases. However, this decline can be problematic, as attackers may continuously reinforce the model’s reliance on backdoor triggers. It is therefore essential to regulate the perturbation strength within an effective range to prevent the model from reverting to decision rules based on backdoor trigger dependencies, while preserving robust generalization.

To address this, we propose adaptive scaling, a mechanism that dynamically adjusts the strength of benign adversarial perturbation based on its real-time influence on the model. Leveraging the adversarial nature of benign adversarial perturbation, we use its training loss value as a proxy for effectiveness: a high loss indicates strong gradient influence, while a decreasing loss suggests that the model is adapting and the defense is losing effectiveness. Adaptive scaling counteracts this by adjusting a scaling factor that amplifies the perturbation magnitude whenever its effect weakens. This mechanism maintains benign adversarial perturbation within a desirable range, ensuring continued suppression of backdoor dependencies throughout training without compromising performance on clean inputs.

Specifically, the scaling factor is governed by three key components. The first is the scaling step size $\delta$, which is a hyperparameter that controls the magnitude of adjustment. The second is the loss ratio $\alpha_t$, measuring the relative change between the current and previous loss values to capture short-term fluctuations. The third is the smoothing factor $\beta$, designed to reduce sensitivity to transient variations and reflect the overall trend of the loss. To compute $\beta$, we use a sliding window that averages recent loss ratios, enabling the mechanism to detect consistent upward or downward trends and stabilize the adjustment of the scaling factor. The complete adaptive scaling procedure \textit{AdaptiveScaling} is detailed in the appendix.

\section{Experiments}

\subsection{Experimental Setup}

\begin{table*}
    \centering
    \caption{Performance of different defenses on non-IID datasets.}
    \label{tab:summary_results_noiid}
    \begin{tabular}{l l | S[table-format=3.2, table-space-text-post=\%] S[table-format=3.2, table-space-text-post=\%]  S[table-format=3.2, table-space-text-post=\%] | S[table-format=3.2, table-space-text-post=\%] S[table-format=3.2, table-space-text-post=\%] S[table-format=3.2, table-space-text-post=\%] | S[table-format=3.2, table-space-text-post=\%] S[table-format=3.2, table-space-text-post=\%] S[table-format=3.2, table-space-text-post=\%]}
        \toprule
        &  & \multicolumn{3}{c|}{ResNet-18 (CIFAR-10)} & \multicolumn{3}{c|}{VGG-19 (CIFAR-10)} & \multicolumn{3}{c}{ResNet-18 (CIFAR-100)} \\
        \cmidrule(lr){3-5} \cmidrule(lr){6-8} \cmidrule(lr){9-11}
        &  & {BadNets} & {LP} & {A3FL} & {BadNets} & {LP} & {A3FL} & {BadNets} & {LP} & {A3FL} \\

        \midrule
                & BBSR & 99.37\% & 93.98\% & 100.00\%  & 75.66\% & 95.91\% & 100.00\%  & 99.53\% & 83.27\% & 100.00\%  \\
        FedAvg  & ABSR & 97.88\% & 91.29\% & 100.00\%  & 60.16\% & 85.06\% & 100.00\%  & 98.50\% & 76.57\% & 100.00\%  \\
                & ACC  & 85.77\% & 75.03\% & 86.18\%   & 78.26\% & 73.10\% & 78.95\%   & 59.39\% & 43.05\% & 58.36\%   \\
        \midrule
                & BBSR & 11.06\% & 97.51\% & 49.57\%  & 100.00\% & 29.01\% & 99.31\%  & 4.39\% & 97.34\% & 5.40\%  \\
        Krum    & ABSR & \textbf{\phantom{0}1.27\%}  & 93.96\% & 10.42\%  & 49.19\%  & \textbf{\phantom{0}2.28\%}  & 31.88\%  & 1.24\% & 64.39\% & 1.31\%  \\
                & ACC  & 44.00\% & 64.37\% & 45.10\%  & 33.38\%  & 47.76\% & 36.10\%  & 24.03\% & 31.16\% & 26.64\%  \\
        \midrule
                  & BBSR & 84.23\% & 87.11\% & 100.00\%  & 59.97\% & 95.97\% & 100.00\%  & 9.30\% & 96.84\% & 100.00\%  \\
        MultiKrum & ABSR & 56.08\% & 20.53\% & 99.71\%  & 22.18\% & 61.82\% & 59.74\%  & 4.23\% & 91.36\% & 99.94\%  \\
                  & ACC  & 78.84\% & 71.28\% & 79.35\%  & 70.06\% & 67.47\% & 70.62\%  & 51.32\% & 38.51\% & 51.03\%  \\
        \midrule
                 & BBSR & 96.24\% & 80.44\% & 100.00\%  & 12.86\% & 14.72\% & 100.00\%  & 48.01\% & 91.41\% & 100.00\%  \\
        FLTrust  & ABSR & 88.57\% & 29.26\% & 100.00\%  & 8.66\% & 6.00\% & 100.00\%  & 64.38\% & 42.52\% & 100.00\%  \\
                 & ACC  & 77.89\% & 67.70\% & 78.63\%  & 63.15\% & 67.66\% & 67.52\%  & 48.01\% & 34.99\% & 48.42\%  \\
        \midrule
                 & BBSR & 96.62\% & 9.56\% & 100.00\%  & 62.57\% & 16.11\% & 100.00\%  & 97.75\% & 35.70\% & 100.00\%  \\
        RLR      & ABSR & 70.32\% & 3.63\% & 97.64\%  & 38.31\% & 6.75\% & 100.00\%  & 93.01\% & 0.94\% & 100.00\%  \\
                 & ACC  & 79.55\% & 65.58\% & 78.86\%  & 62.57\% & 63.21\% & 63.52\%  & 46.50\% & 35.70\% & 47.21\%  \\
        \midrule
                 & BBSR & 98.77\% & 97.62\% & 100.00\%  & 91.08\% & 98.09\% & 100.00\%  & 49.07\% & 91.85\% & \textbf{\phantom{0}2.93\%}  \\
        FLAME    & ABSR & 93.39\% & 95.61\% & 100.00\%  & 79.17\% & 71.49\% & 100.00\%  & \textbf{\phantom{0}0.56\%} & 86.69\% & \textbf{\phantom{0}1.26\%}  \\
                 & ACC  & 78.00\% & 68.77\% & 79.46\%  & 64.68\% & 66.29\% & 69.68\%  & 49.07\% & 36.68\% & 48.19\%  \\
        \midrule
                 & BBSR & 15.83\% & 36.89\% & 100.00\%  & 9.97\% & 11.14\% & 100.00\%  & 82.45\% & 20.64\% & 100.00\%  \\
        FLIP     & ABSR & 4.37\% & 8.60\% & 100.00\%  & 2.48\% & 3.32\% & 100.00\%  & 5.92\% & 4.77\% & 100.00\%  \\
                 & ACC  & 85.95\% & 74.67\% & 85.75\%  & 79.17\% & 73.87\% & 78.38\%  & 57.41\% & 43.56\% & 58.23\%  \\  
        \midrule
                 & BBSR & \textbf{\phantom{0}2.51\%} & \textbf{\phantom{0}4.22\%} & \textbf{\phantom{0}5.89\%}  & \textbf{\phantom{0}5.78\%} & \textbf{\phantom{0}5.50\%} & \textbf{12.13\%}  & \textbf{\phantom{0}1.11\%} & \textbf{\phantom{0}0.96\%} & 4.48\%  \\
        FedBAP (Ours)   & ABSR & 1.55\% & \textbf{\phantom{0}2.26\%} & \textbf{\phantom{0}2.40\%}  & \textbf{\phantom{0}2.26\%} & 2.84\% & \textbf{\phantom{0}2.78\%}  & 0.58\% & \textbf{\phantom{0}0.38\%} & 2.55\%  \\
                 & ACC  & \textbf{89.98\%} & \textbf{\phantom{0}78.64\%} & \textbf{89.29\%}  & \textbf{87.24\%} & \textbf{78.40\%} & \textbf{86.32\%}  & \textbf{63.12\%} & \textbf{47.22\%} & \textbf{63.05\%}  \\
        
        \bottomrule
    \end{tabular}
\end{table*}

\begin{figure*}[t]
    \centering
    \begin{subfigure}[b]{0.24\linewidth}
        \centering
        \includegraphics[width=\linewidth]{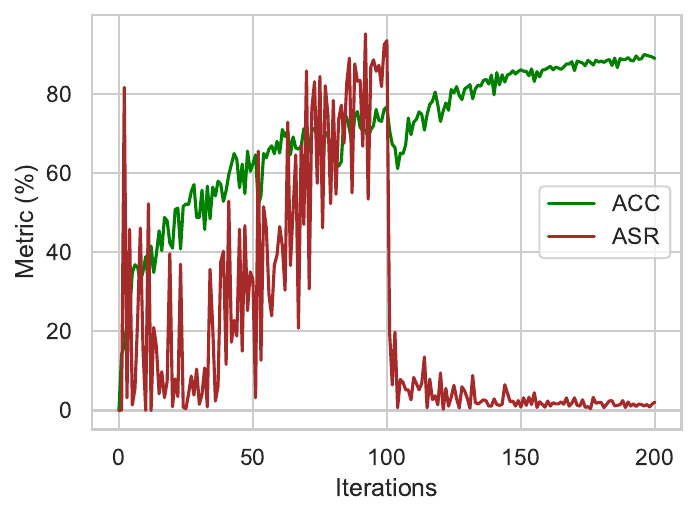}
        \caption{FedBAP (Ours)}
    \end{subfigure}
    \hfill
    \begin{subfigure}[b]{0.24\linewidth}
        \centering
        \includegraphics[width=\linewidth]{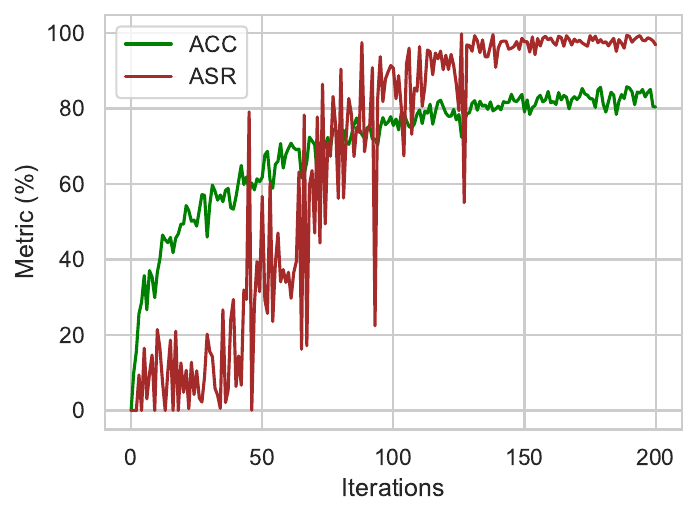}
        \caption{FedAvg}
    \end{subfigure}
    \hfill
    \begin{subfigure}[b]{0.24\linewidth}
        \centering
        \includegraphics[width=\linewidth]{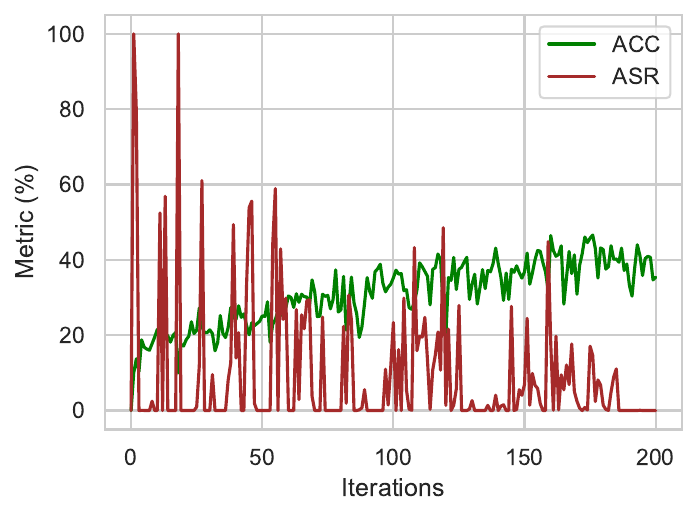}
        \caption{Krum}
    \end{subfigure}
    \hfill
    \begin{subfigure}[b]{0.24\linewidth}
        \centering
        \includegraphics[width=\linewidth]{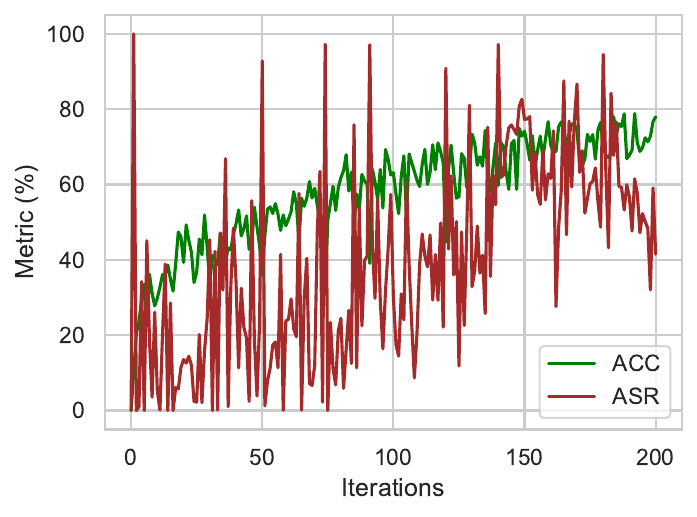}
        \caption{MultiKrum}
    \end{subfigure}

    \par\medskip

    \begin{subfigure}[b]{0.24\linewidth}
        \centering
        \includegraphics[width=\linewidth]{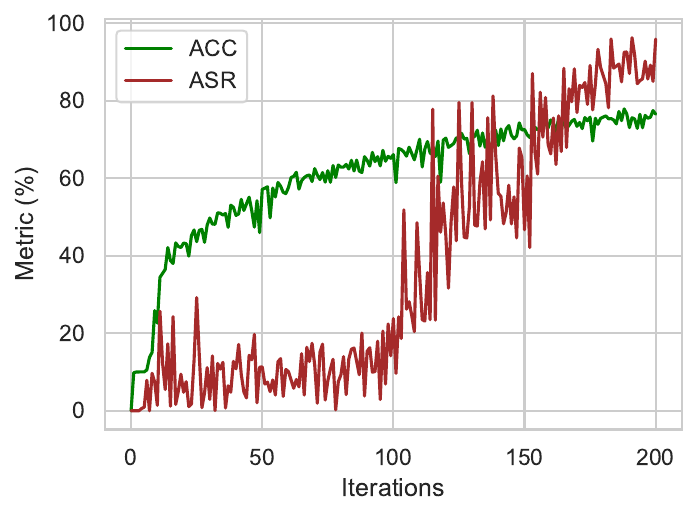}
        \caption{FLTrust}
    \end{subfigure}
    \hfill
    \begin{subfigure}[b]{0.24\linewidth}
        \centering
        \includegraphics[width=\linewidth]{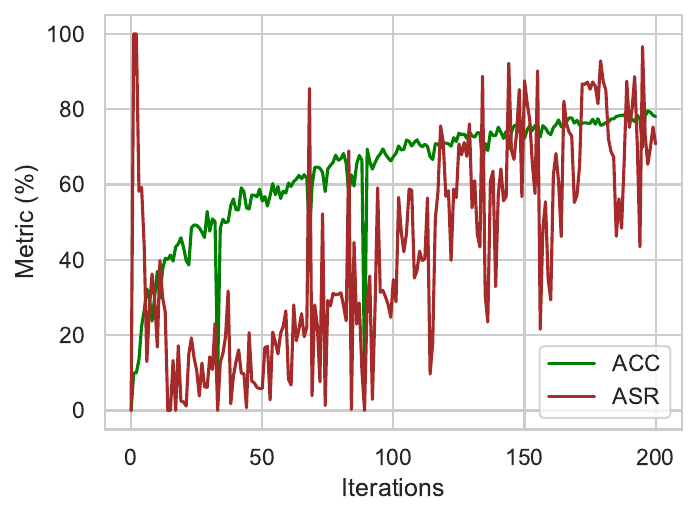}
        \caption{RLR}
    \end{subfigure}
    \hfill
    \begin{subfigure}[b]{0.24\linewidth}
        \centering
        \includegraphics[width=\linewidth]{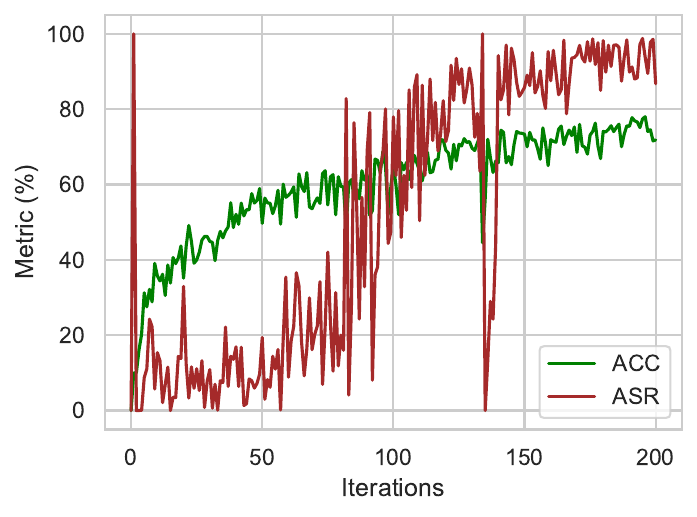}
        \caption{FLAME}
    \end{subfigure}
    \hfill
    \begin{subfigure}[b]{0.24\linewidth}
        \centering
        \includegraphics[width=\linewidth]{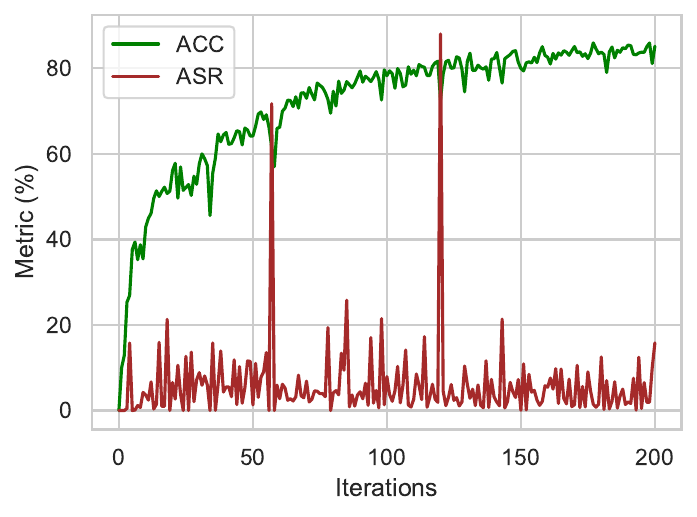}
        \caption{FLIP}
    \end{subfigure}
    \caption{ResNet-18 trained with different defenses on the non-IID CIFAR10 dataset against the BadNets attack.}
    \Description{}
    \label{badnets}
\end{figure*}

\textbf{FL Settings.}
We consider two widely-used benchmark datasets CIFAR-10 and CIFAR-100 to evaluate FedBAP. We use the method in \cite{fang2020local} to distribute the training images to the clients. For CIFAR-10, we employ the commonly used ResNet-18 and VGG-19 architectures as the global model, while for CIFAR-100, we adopt ResNet-18 as the global model. By default, our federated training setup consists of 100 clients, 10\% of which are malicious. In each training round, 10\% of clients are randomly selected to perform local model training. We assume a non-IID data distribution with a concentration parameter $h$ of 0.9, following previous works \cite{lyu2023poisoning, fang2023vulnerability, zhang2022neurotoxin}. The FL training process consists of 200 communication rounds, with each selected client training the local model for 2 epochs. The start round for FedBAP is set to 100, and unless otherwise specified, the scaling step size $\delta$ is set to 1. More setup details can be found in the appendix.

\textbf{Baselines.}
We consider three types of state-of-the-art (SOTA) targeted attacks: BadNets ~\cite{gu2017badnets}, LP ~\cite{zhuang2023backdoor}, and A3FL ~\cite{zhang2023a3fl}. The performance of FedBAP is compared with six SOTA defenses: Krum ~\cite{blanchard2017machine}, MultiKrum, FLTrust ~\cite{cao2020fltrust}, RLR ~\cite{ozdayi2021defending}, FLAME ~\cite{nguyen2022flame}, and FLIP ~\cite{zhang2022flip} . Detailed descriptions of these comparison schemes can be found in the appendix.

\textbf{Evaluation Metrics.}
We consider Average Backdoor Success Rate (ABSR), Best Backdoor Success Rate (BBSR), and Main Task Accuracy (ACC) as evaluation metrics to assess the effectiveness of FedBAP. ABSR represents the average proportion of backdoor samples misclassified as the attack target label over the last 20 rounds. It reflects the effectiveness of the defense. BBSR denotes the highest proportion of backdoor samples misclassified as the attack target label within the last 20 rounds. It reflects the stability of the defense. ACC measures the accuracy on benign samples, representing the proportion of correctly classified benign inputs. It reflects the usability of the defense.

\subsection{Overall Performance}

Table \ref{tab:summary_results_noiid} presents the performance of different defenses against three backdoor attacks across various model architectures and datasets in non-IID settings, while the results for IID settings are provided in the appendix. Figure \ref{badnets} illustrates the backdoor success rates (BSR) and ACC curves of different defenses against the BadNets attack on ResNet-18 with CIFAR-10, while the results for the LP and A3FL attacks are shown in the appendix. 

As shown in Table \ref{tab:summary_results_noiid}, FedBAP consistently achieves the lowest ABSR across most attack scenarios, highlighting its strong effectiveness in mitigating diverse backdoor threats. In terms of stability, FedBAP maintains the lowest BBSR across most cases, with minimal fluctuations in BSR over time. The BSR curves in Figure \ref{badnets} show that FedBAP remains stable and does not exhibit sudden spikes, which is critical in ensuring reliable defense performance. Regarding usability, FedBAP consistently achieves the highest ACC across all attack scenarios, outperforming other baselines. This can be attributed to the nature of benign adversarial perturbations, which serve as a form of robust training that improves generalization and enhances the model's resilience to abnormal or noisy data.

The results consistently demonstrate that FedBAP outperforms existing defenses by effectively mitigating backdoor attacks while maintaining high main task accuracy. Its ability to provide stable and robust defense in both IID and non-IID settings highlights its practical applicability in FL scenarios. These findings confirm that FedBAP is a highly effective and generalizable defense against various types of backdoor attacks.

\subsection{Ablation Study}

We conduct ablation studies on the CIFAR-10 dataset under the BadNets attack to evaluate the contribution of each component of FedBAP, including Benign Adversarial Perturbation (BAP), Adaptive Scaling (AS), and Perturbation Pattern Generation (PG). Notably, in the w/o PG setting, the learned perturbation pattern is replaced with a randomly initialized one instead of executing PatternGen. As shown in Table~\ref{tab:performance_metrics}, removing any of these components results in a significant decline in defense performance, indicating that each module plays an essential role in the effectiveness of the overall framework.

\begin{table}
  \caption{Ablation study.}
  \label{tab:performance_metrics}
  \begin{tabular}{c|ccccc}
    \toprule
    & w/o BAP & w/o AS \& PG & w/o PG & w/o AS & FedBAP \\
    \midrule
    BBSR & 99.37 & 99.17 & 5.32 & 98.97 & \textbf{2.51} \\
    ABSR & 97.88 & 97.92 & 2.25 & 97.28 & \textbf{1.55} \\
    ACC  & 85.77 & 88.61 & 89.15 & 89.5 & \textbf{89.98} \\
    \bottomrule
  \end{tabular}
\end{table}

\subsection{Impact of the Proportion of Malicious Clients}

\begin{figure}[t]
    \centering
    \begin{subfigure}[b]{0.49\linewidth}
        \centering
        \includegraphics[width=\linewidth]{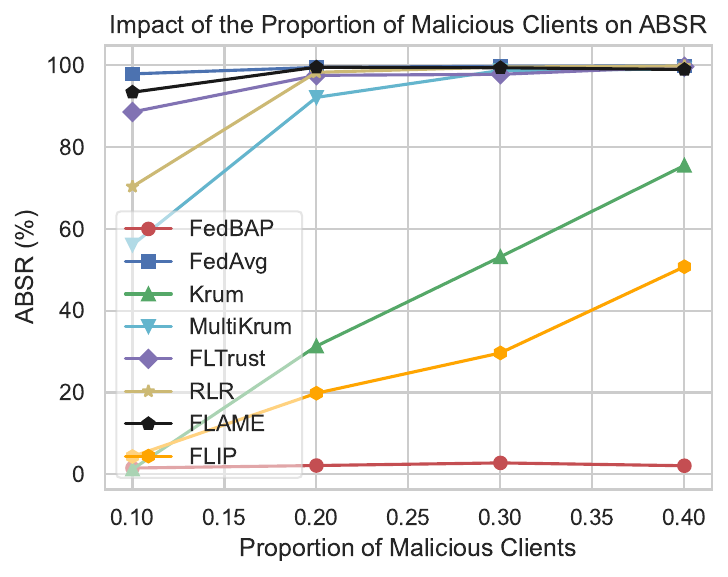}
        \caption{ABSR}
    \end{subfigure}
    \begin{subfigure}[b]{0.49\linewidth}
        \centering
        \includegraphics[width=\linewidth]{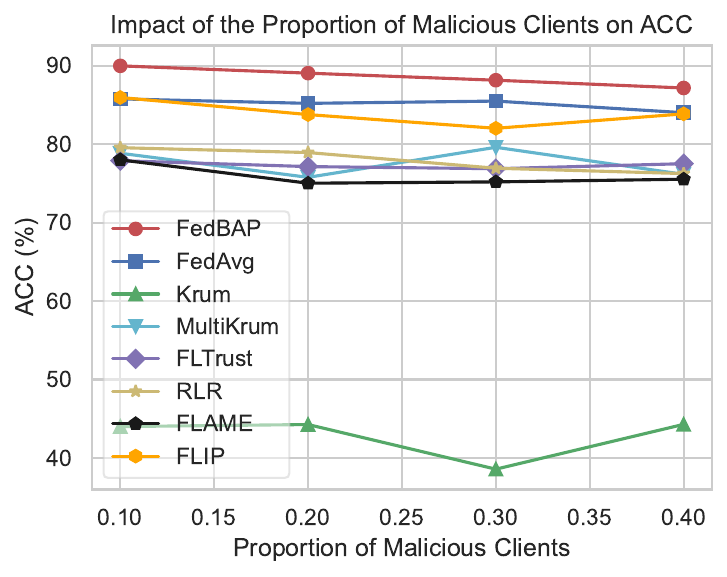}
        \caption{ACC}
    \end{subfigure}
    \caption{Impact of the malicious clients proportion on different defenses under the BadNets attack.}
    \Description{}
    \label{proportion_badnet}
\end{figure}

\begin{figure}[t]
    \centering
    \begin{subfigure}[b]{0.49\linewidth}
        \centering
        \includegraphics[width=\linewidth]{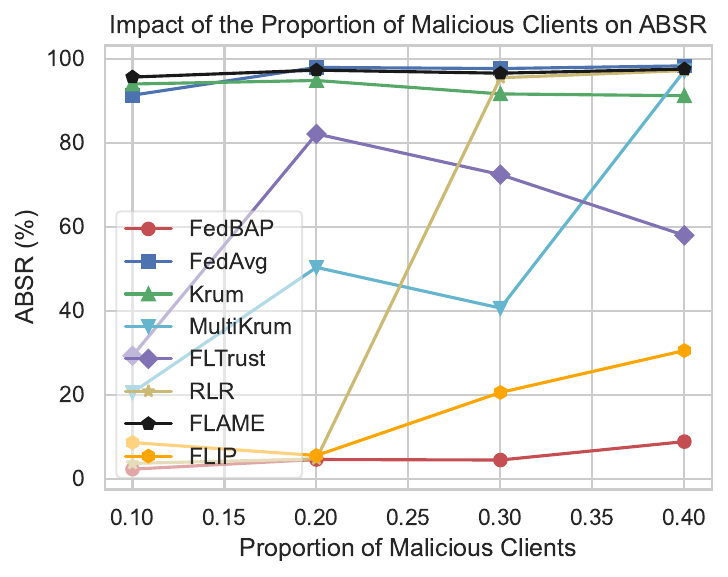}
        \caption{ABSR}
    \end{subfigure}
    \begin{subfigure}[b]{0.49\linewidth}
        \centering
        \includegraphics[width=\linewidth]{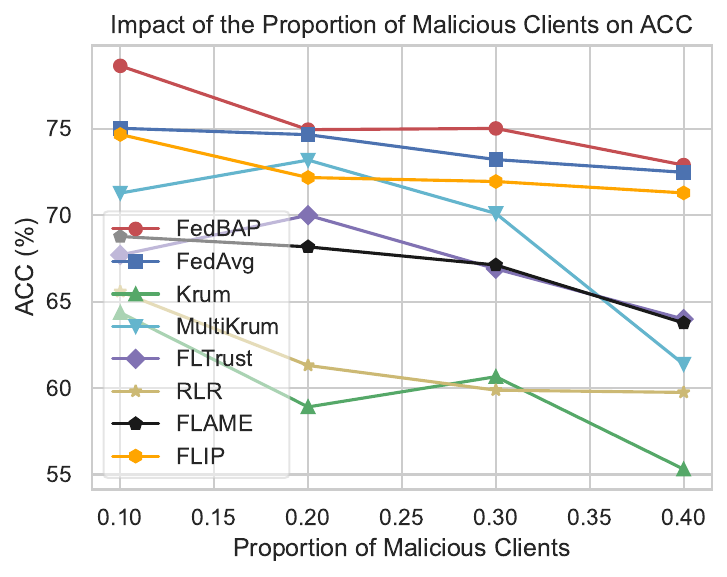}
        \caption{ACC}
    \end{subfigure}
    \caption{Impact of the malicious clients proportion on different defenses under LP attack.}
    \Description{}
    \label{proportion_lp}
\end{figure}

\begin{figure}[t]
    \centering
    \begin{subfigure}[b]{0.49\linewidth}
        \centering
        \includegraphics[width=\linewidth]{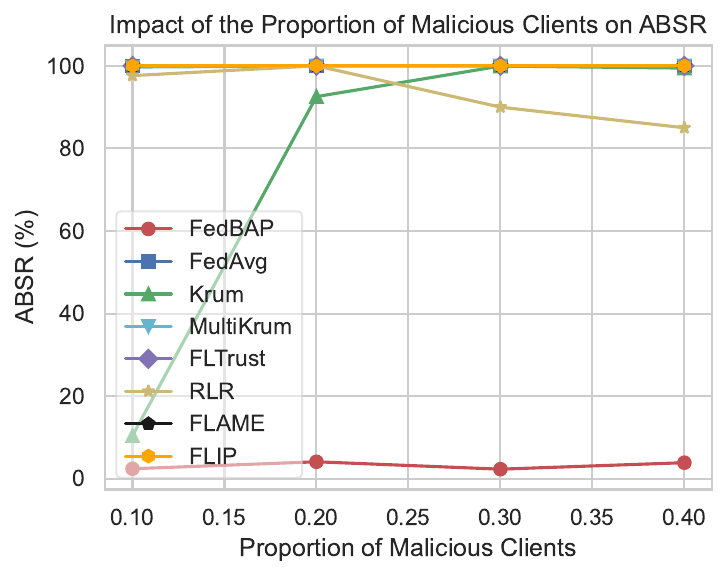}
        \caption{ABSR}
    \end{subfigure}
    \begin{subfigure}[b]{0.49\linewidth}
        \centering
        \includegraphics[width=\linewidth]{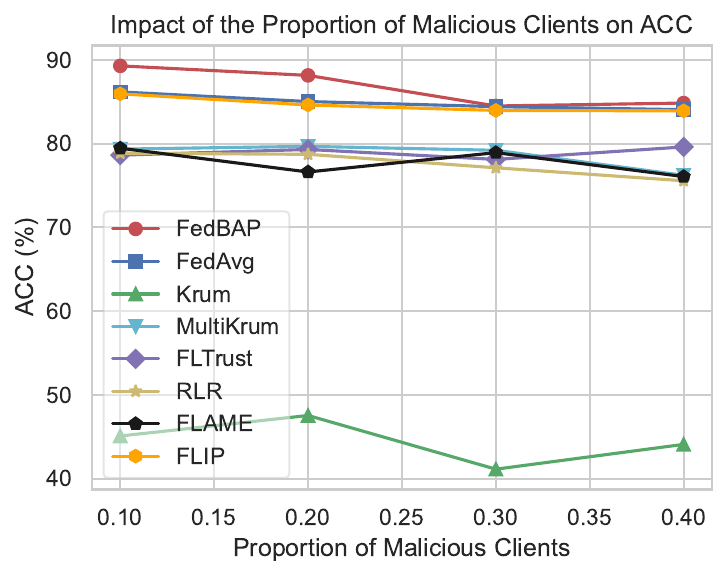}
        \caption{ACC}
    \end{subfigure}
    \caption{Impact of the malicious clients proportion on different defenses under A3FL attack.}
    \Description{}
    \label{proportion_a3fl}
\end{figure}

We evaluate the impact of the proportion of malicious clients using the CIFAR-10 dataset. Specifically, we vary the proportion of malicious clients from 0.1 to 0.4 under three representative attacks: BadNets, LP, and A3FL. It is worth noting that we increase the $\delta$ of FedBAP to 1.5 under LP and to 3 under A3FL, as these are more sophisticated attack strategies. Figures \ref{proportion_badnet}, \ref{proportion_lp}, and \ref{proportion_a3fl} respectively illustrate the ABSR and ACC trends of various defenses under the BadNets, LP, and A3FL attacks. The results show that increasing the proportion of malicious clients leads to a significant drop in the performance of most defense methods, except FedBAP. FedBAP consistently achieves the lowest ABSR across all settings, staying below 6\% under LP and A3FL, and under 3\% for BadNets. It also maintains higher ACC than other defenses, demonstrating strong robustness across different attack scenarios. This is because FedBAP encourages the model to learn robust features rather than overfitting to malicious patterns, thus maintaining stability even under stronger attack intensities.

\subsection{Impact of the non-IIDness}

\begin{figure}[t]
    \centering
    \begin{subfigure}[b]{0.49\linewidth}
        \centering
        \includegraphics[width=\linewidth]{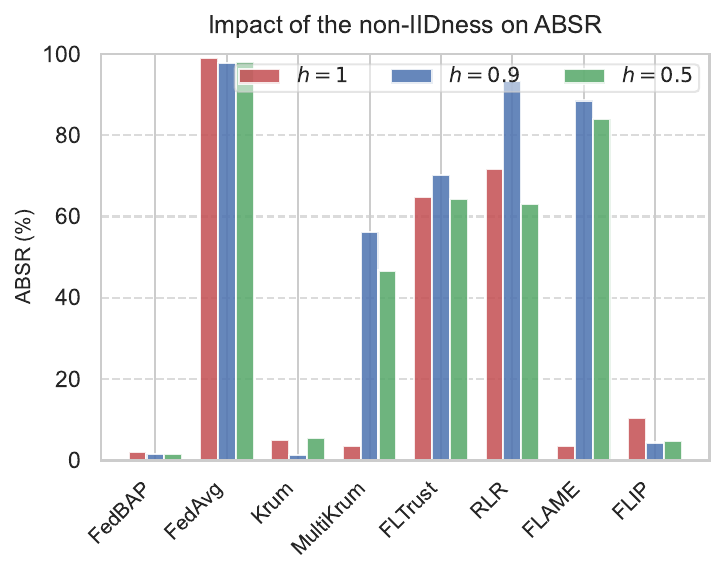}
        \caption{ABSR}
    \end{subfigure}
    \begin{subfigure}[b]{0.49\linewidth}
        \centering
        \includegraphics[width=\linewidth]{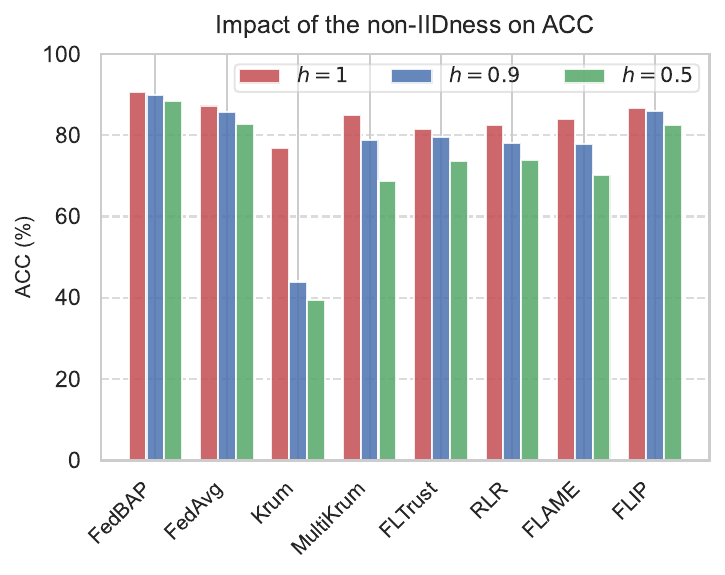}
        \caption{ACC}
    \end{subfigure}
    \caption{Impact of the non-IIDness on different defenses under BadNets attack.}
    \Description{}
    \label{IIDness}
\end{figure}

To investigate the effect of data heterogeneity, we evaluate all methods under both IID and non-IID settings, where $h=1.0$ corresponds to the IID setting and smaller $h$ values (e.g., $h=0.9$ and $h=0.5$) indicate increasing levels of non-IID distribution. We conduct experiments on the CIFAR-10 dataset under the BadNets attack. As shown in Figure \ref{IIDness}, most baseline methods suffer significant degradation in ABSR and ACC as the degree of non-IIDness increases. In contrast, FedBAP consistently maintains low attack success rates and exhibits only a minor drop in ACC under non-IID settings. This demonstrates that FedBAP is not only robust against backdoor attacks, but also resilient to performance degradation caused by data heterogeneity.

\subsection{Impact of the Scaling Step Size}

\begin{figure}[h]
  \centering
  \includegraphics[width=0.8\linewidth]{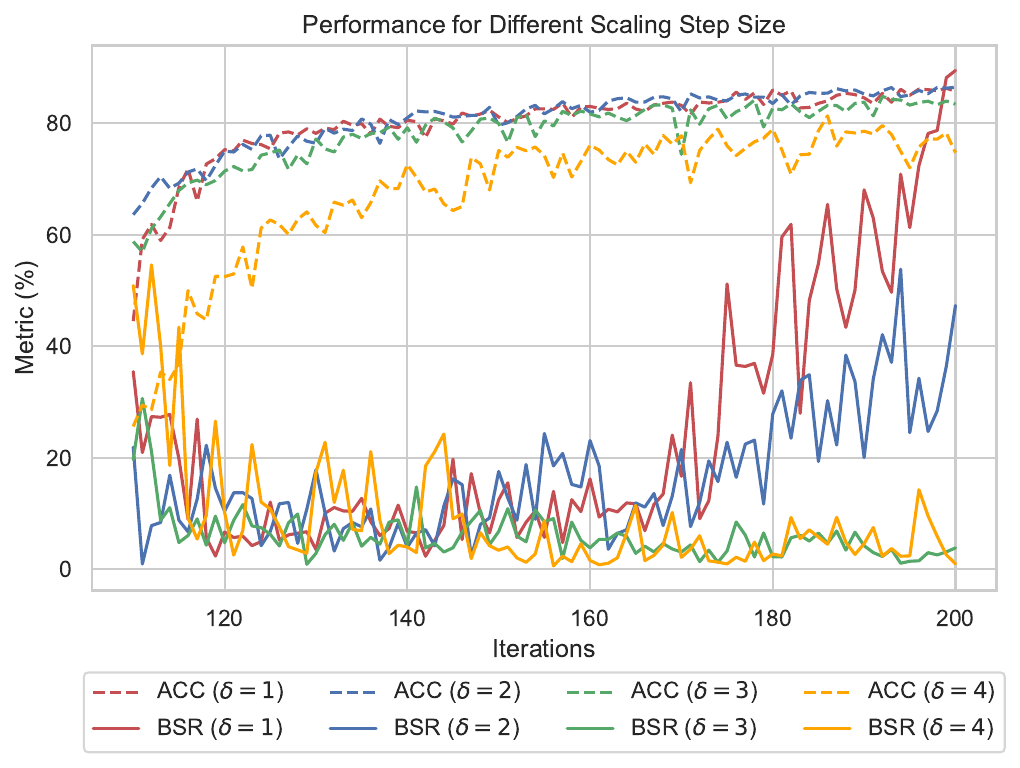}
  \caption{Impact of the scaling step size.}
  \Description{}
  \label{delta}
\end{figure}

To evaluate the impact of the scaling step size $\delta$ on the effectiveness of our defense, we compare the ACC and BSR across different values of $\delta$ using the CIFAR-10 dataset under A3FL attacks, where the proportion of malicious clients is set to 40\%. As shown in Figure \ref{delta}, we present results starting from round 110, as the defense mechanism is activated at round 100. Experimental results show that increasing $\delta$ effectively enhances backdoor suppression but may also degrade model performance. When $\delta < 3$, the backdoor effect tends to re-emerge during training. In contrast, BSR drops below 5\% and remains suppressed when $\delta = 3$. However, further increasing $\delta$ to 4 leads to a noticeable decline in ACC, indicating potential instability. These results suggest that a small benign adversarial perturbation may be insufficient to eliminate the backdoor, while an overly large one can harm the model’s performance. Therefore, choosing an appropriate $\delta$ is crucial for balancing robustness and utility. Among the evaluated settings, $\delta = 3$ strikes the best trade-off, offering effective backdoor suppression while maintaining stable accuracy on the main task.

\section{Conclusion}

In this paper, we propose FedBAP, a robust FL defense mechanism against backdoor attacks. In FedBAP, we propose a novel defense approach that stems from the behavior of model training itself, focusing on the inherent reliance of the model on backdoor features.
FedBAP utilizes perturbation trigger to generate benign adversarial perturbations and employs adaptive scaling to control the perturbation magnitude, thereby mitigating the influence of malicious model updates while preserving overall model accuracy. This method of weakening reliance from the source, guiding training locally, and stabilizing the system through dynamic adjustment ensures both enhanced defense effectiveness and the preservation of model accuracy. Our extensive evaluation on two benchmark datasets, three backdoor attack strategies, and six defense methods demonstrates that FedBAP effectively suppresses backdoor effects and achieves SOTA performance, consistently outperforming existing defense approaches in multiple scenarios.
In future work, we plan to extend our method to asynchronous FL and personalized FL to address backdoor attacks faced in these FL scenarios.

\begin{acks}
This work was supported by the National Key Research and Development Program of China under Grant 2022YFB3104502, in part by the National Natural Science Foundation of China under Grant 62441237, Grant U24A20336, Grant 62272348, Grant 62202197, the Key Research and Development Program of Wuhan under Grant 2024050702030090, Wuhan Science and Technology Joint Project for Building a Strong Transportation Country under Grant 2023-2-7, and Wuhan City Joint Innovation Laboratory for Next-Generation Wireless Communication Industry Featuring Satellite-Terrestrial Integration under Grant 4050902040448.
\end{acks}

\bibliographystyle{ACM-Reference-Format}
\bibliography{sample-base}

\appendix
\newpage

\section{Algorithms}

\begin{algorithm}[h]
\caption{Perturbation Trigger Mask Generation}
\label{alg:algorithm2}
\raggedright
\textbf{MaskGen:} \\
\hspace*{1em}\textbf{Input:} Global model $w_t$.\\
\hspace*{1em}\textbf{Parameter:} Class numbers $N_{class}$, epochs $E_2$, learning rate $\eta_2$, backdoor accuracy threshold $acc_{TH}$, regularization coefficient $\lambda$.\\
\hspace*{1em}\textbf{Output:} Trigger mask $M$.
\begin{algorithmic}[1]
\STATE $S \gets $ the set of all clients
\FOR{$i \in S$ \textbf{in parallel}}
\FOR{$y_{t}$ = 0 to $N_{class}$}
\STATE init $M_{i}^{y_t}$ and $\Delta x_{i}^{y_t} $ randomly
\FOR{$e$ = 1 to $E_2$}
\FOR{$(x, y) \in \mathcal{D}_i$}
\STATE $loss \gets \ell\left(w_t(A(x, M_{i}^{y_t}, \Delta x_{i}^{y_t})), y_{t}\right) + \lambda \cdot \| M_{i}^{y_t} \| $
\STATE $M_{i}^{y_t} \gets M_{i}^{y_t} - \eta_2 \nabla_{M_{i}^{y_t}} loss$
\STATE $\Delta x_{i}^{y_t} \gets \Delta x_{i}^{y_t} - \eta_2 \nabla_{\Delta x_{i}^{y_t}} loss$
\ENDFOR 
\STATE compute backdoor accuracy $acc$
\IF{$acc \geq acc_{TH}$ and $\| M_{i}^{y_t} \| <  \| M_{i}^{best} \|$}
\STATE $M_{i}^{best} \gets M_{i}^{y_t}$
\ENDIF
\ENDFOR
\ENDFOR
\ENDFOR
\FOR{$j=0$ to $row_{M}$}
\FOR{$k=0$ to $col_{M}$}
\STATE $M[j][k] \gets \mathbb{I}\left\{ \frac{1}{|S|} \sum_{i \in S} M_i^{best}[j][k] \geq 0.5 \right\}$
\ENDFOR
\ENDFOR
\STATE \textbf{return} $M$
\end{algorithmic}
\end{algorithm}

\begin{algorithm}[h]
\caption{Perturbation Trigger Pattern Generation}
\label{alg:algorithm3}
\raggedright
\textbf{PatternGen:} \\
\hspace*{1em}\textbf{Input:} Global model $w_t$, trigger mask $M$.\\
\hspace*{1em}\textbf{Parameter:} Epochs $E_3$, learning rate $\eta_3$. \\
\hspace*{1em}\textbf{Output:} Trigger patterns $\Delta x$.
\begin{algorithmic}[1]
\STATE $S \gets $ the set of all clients
\FOR{$i \in S$ \textbf{in parallel}}
\STATE init $\Delta x_i$ randomly
\FOR{$e$ = 1 to $E_3$}
\FOR{$(x, y) \in \mathcal{D}_i$}
\STATE $r_1 \gets w_{t}^{penultimate}(A(x, M, \Delta x_i))$
\STATE $r_2 \gets w_{t}^{penultimate}(x)$
\STATE $\Delta x_i \gets \Delta x_i - \eta_3 \nabla_{\Delta x_i} \frac{r_1 \cdot r_2}{\|r_1\| \|r_2\|}$
\ENDFOR 
\ENDFOR
\ENDFOR
\STATE \textbf{return} $\Delta x$
\end{algorithmic}
\end{algorithm}

\begin{algorithm}[h]
\caption{Benign Adversarial Perturbation Generation}
\label{alg:algorithm4}
\raggedright
\textbf{BAPGen:} \\
\hspace*{1em}\textbf{Input:} Scaling factor $c_t$, local model $w^\prime$, client $i$, trigger mask $M$, perturbation trigger $\Delta x_{i}$.\\
\hspace*{1em}\textbf{Parameter}: Benign adversarial perturbation epochs $E_4$, learning rate $\eta_4$.\\
\hspace*{1em}\textbf{Output:} Benign adversarial perturbation loss value $loss_{t}^i$, local model $w^\prime$.
\begin{algorithmic}[1]
\FOR{$e$ = 1 to $E_4$}
\FOR{$(x, y) \in \mathcal{D}_i$}
\STATE $loss \gets c_t \cdot \ell(w^\prime(A(x, M, \Delta x_i), y)$
\STATE $w^\prime \gets w^\prime - \eta_4 \nabla_{w^\prime} loss$
\ENDFOR
\ENDFOR
\STATE $loss_{t}^i \gets$ the average value of $loss$ over $E_4$ epochs on $\mathcal{D}_i$
\STATE \textbf{return} $loss_{t}^i$, $w^\prime$
\end{algorithmic}
\end{algorithm}

\begin{algorithm}[h]
\caption{Adaptive Scaling}
\label{alg:algorithm5}
\raggedright
\textbf{AdaptiveScaling:} \\
\hspace*{1em}\textbf{Input:} Benign adversarial perturbation loss value $loss_{t}$, scaling factor $c_t$, round $t$.\\
\hspace*{1em}\textbf{Parameter:} Scaling step size $\delta$, start round $t_s$, window size $k$.\\
\hspace*{1em}\textbf{Output:} Updated scaling factor $c_{t+1}$.
\begin{algorithmic}[1]
\STATE compute $\alpha_t \gets 
\begin{cases}
1, & \text{if } t = t_s\\
\frac{\mathrm{loss}_t}{\mathrm{loss}_{t - 1}}, & \text{if } t>t_s
\end{cases}$
\STATE compute $\beta \gets 
\begin{cases}
1, & \text{if } t - t_s = 0\\
\frac{\sum_{j = ts}^{t - 1}\alpha_j}{t - ts}, & \text{if } 0 < t - t_s < k\\
\frac{\sum_{j = t - k}^{t - 1}\alpha_j}{k}, & \text{if } t - t_s \geq k
\end{cases}$
\STATE compute $c_{t + 1} \gets  
\begin{cases}
c_t+\frac{\delta\beta}{\sqrt{\alpha_t}}, & \text{if } \alpha_t \geq 1\\
c_t+\frac{\delta\beta}{\alpha_t^2}, & \text{if } \alpha_t < 1
\end{cases}$
\STATE \textbf{return} $c_{t+1}$
\end{algorithmic}
\end{algorithm}

We present the detailed workflow of four key algorithms in FedBAP, namely \textit{MaskGen}, \textit{PatternGen}, \textit{BAPGen}, and \textit{AdaptiveScaling}. These algorithms constitute the core of our defense mechanism and are essential for enhancing the robustness and generalization of the global model. Their pseudocode is presented in Algorithms~\ref{alg:algorithm2}–\ref{alg:algorithm5} to facilitate reproducibility and a better understanding of our approach.

\section{Details of Experiment Setup}
All experiments are conducted using the PyTorch framework on an NVIDIA RTX 3080 Ti GPU. The implementations of backdoor attacks, including BadNets ~\cite{gu2017badnets} and LP ~\cite{zhuang2023backdoor}, as well as defenses such as Krum ~\cite{blanchard2017machine}, MultiKrum, FLTrust ~\cite{cao2020fltrust}, RLR ~\cite{ozdayi2021defending}, FLAME ~\cite{nguyen2022flame}, and FLIP ~\cite{zhang2022flip}, were directly used from implementations by LP ~\cite{zhuang2023backdoor}. The A3FL ~\cite{zhang2023a3fl} attack is re-implemented for fair comparison, with reference to the original code. The detailed experimental hyperparameter settings are shown in Tables \ref{tab:parameters1} and \ref{tab:parameters2}.

\begin{table}[H]
  \caption{Hyper-parameters settings in FL.}
  \label{tab:parameters1}
  \begin{tabular}{cc}
    \toprule
    \textbf{Experimental parameters} & \textbf{Parameter settings} \\
    \midrule
    Number of clients & 100 \\
    Select clients proportion & 0.1 \\
    Malicious clients proportion & 0.1 \\
    Malicious data proportion & 0.5 \\
    Trigger size & $5 \times 5$ \\
    Global training rounds & 200 \\
    Local epochs & 2 \\
    Learning rate & 0.01 \\
    Batch size & 64 \\
    \bottomrule
  \end{tabular}
\end{table}

\begin{table}[H]
  \caption{Hyper-parameters settings in FedBAP.}
  \label{tab:parameters2}
  \begin{tabular}{ccc}
    \toprule
      \,  & \textbf{Experimental parameters} & \textbf{Parameter settings} \\
    \midrule
    $E_1$ & Local epochs & 2\\
    $\eta_1$ & Learning rate & 0.01\\
    $t_s$ & Start round & 100\\
    $E_2$ & Local epochs for \textit{MaskGen} & 100\\
    $\eta_2$ & Learning rate for \textit{MaskGen} & 0.1\\
    $acc_{TH}$ & Backdoor accuracy threshold & 0.9\\
    $\lambda$ & Weight for backdoor distance & 0.01\\
    $E_3$ & Local Epochs for \textit{PatternGen} & 100\\
    $\eta_3$ & Learning rate for \textit{PatternGen} & 10\\
    $E_4$ & Local epochs for \textit{BAPGen} & 10\\
    $\eta_4$ & Learning rate for \textit{BAPGen} & 0.01\\
    $\delta$ & Scaling step size & 1\\
    $k$ & Window size & 5\\
    \bottomrule
  \end{tabular}
\end{table}

\section{Baseline Attacks}

\begin{itemize}
    \item \textbf{BadNets} \cite{gu2017badnets} is one of the earliest and most influential backdoor attack methods. It demonstrates that deep neural networks trained in an outsourced or untrusted environment can be embedded with malicious behavior. A BadNet maintains high accuracy on clean data but misclassifies specific attacker-chosen inputs containing a trigger. The attack is stealthy and persistent—even retraining the model may not completely remove the backdoor. This foundational work highlights the vulnerability of DNNs to covert manipulation during training.

    \item \textbf{LP} \cite{zhuang2023backdoor} is a novel backdoor attack strategy by identifying backdoor-critical (BC) layers—a small set of layers primarily responsible for model vulnerability. By targeting only these layers, LP achieves comparable attack effectiveness to full-model attacks while significantly improving stealthiness against existing defenses. 

    \item \textbf{A3FL} \cite{zhang2023a3fl} is an adaptive backdoor attack that dynamically adjusts the trigger to align with the global model training dynamics in FL. Unlike traditional static triggers, A3FL optimizes the trigger to survive scenarios where the global model actively tries to unlearn it. It remains effective even under limited attack budgets and demonstrates high attack success rates against a broad range of state-of-the-art defenses.

\end{itemize}

\begin{table*}
    \centering
    \caption{Performance of different defenses on IID datasets.}
    \label{tab:summary_results_iid}
    \begin{tabular}{l l | S[table-format=3.2, table-space-text-post=\%] S[table-format=3.2, table-space-text-post=\%]  S[table-format=3.2, table-space-text-post=\%] | S[table-format=3.2, table-space-text-post=\%] S[table-format=3.2, table-space-text-post=\%] S[table-format=3.2, table-space-text-post=\%] | S[table-format=3.2, table-space-text-post=\%] S[table-format=3.2, table-space-text-post=\%] S[table-format=3.2, table-space-text-post=\%]}
        \toprule
        &  & \multicolumn{3}{c|}{ResNet-18 (CIFAR-10)} & \multicolumn{3}{c|}{VGG-19 (CIFAR-10)} & \multicolumn{3}{c}{ResNet-18 (CIFAR-100)} \\
        \cmidrule(lr){3-5} \cmidrule(lr){6-8} \cmidrule(lr){9-11}
        &  & {BadNets} & {LP} & {A3FL} & {BadNets} & {LP} & {A3FL} & {BadNets} & {LP} & {A3FL} \\

        \midrule
                 & BBSR & 99.71\% & 95.64\% & 100.00\%  & 91.87\% & 95.43\% & 100.00\%  & 99.60\% & 90.76\% & 100.00\%  \\
        FedAvg   & ABSR & 99.11\% & 95.54\% & 100.00\%  & 83.86\% & 94.03\% & 100.00\%  & 98.20\% & 88.95\% & 100.00\%  \\
                 & ACC  & 87.34\% & 74.42\% & 87.41\%  & 82.34\% & 75.69\% & 82.33\%  & 60.17\% & 43.08\% & 60.04\%  \\
        \midrule
                & BBSR & 13.36\% & 98.83\% & 13.40\%  & 10.20\% & 99.57\% & 7.94\%  & 3.10\% & 96.17\% & \textbf{\phantom{0}1.69\%}  \\
        Krum    & ABSR & 4.94\% & 50.86\% & 5.71\%  & 4.37\% & 27.46\% & 3.20\%  & 0.65\% & 89.96\% & \textbf{\phantom{0}0.63\%}  \\
                & ACC  & 76.92\% & 68.77\% & 77.21\%  & 70.36\% & 65.46\% & 69.62\%  & 37.53\% & 29.43\% & 34.56\%  \\
        \midrule
                  & BBSR & 8.44\% & 97.69\% & 10.91\%  & 6.14\% & 94.69\% & 5.75\%  & 1.23\% & 94.28\% & 2.49\%  \\
        MultiKrum & ABSR & 3.51\% & 97.07\% & 4.87\%  & 3.47\% & 55.86\% & 2.73\%  & \textbf{\phantom{0}0.41\%} & 91.07\% & 0.81\%  \\
                  & ACC  & 84.88\% & 73.21\% & 84.02\%  & 79.13\% & 72.97\% & 79.10\%  & 52.77\% & 38.25\% & 51.83\%  \\
        \midrule
                 & BBSR & 78.27\% & 99.94\% & 100.00\%  & 9.81\% & 96.71\% & 100.00\%  & 91.98\% & 97.92\% & 100.00\%  \\
        FLTrust  & ABSR & 64.73\% & 94.37\% & 100.00\%  & 7.82\% & 93.58\% & 100.00\%  & 85.73\% & 86.93\% & 100.00\%  \\
                 & ACC  & 81.61\% & 73.51\% & 82.82\%  & 70.36\% & 71.81\% & 71.67\%  & 50.21\% & 38.75\% & 49.54\%  \\
        \midrule
                 & BBSR & 87.57\% & 34.99\% & 100.00\%  & 51.16\% & 91.36\% & 100.00\%  & 89.74\% & 26.98\% & 100.00\%  \\
        RLR      & ABSR & 71.63\% & 13.96\% & 100.00\%  & 36.46\% & 87.80\% & 100.00\%  & 74.25\% & 12.83\% & 100.00\%  \\
                 & ACC  & 82.56\% & 69.98\% & 82.04\%  & 68.65\% & 68.66\% & 69.77\%  & 48.50\% & 36.91\% & 47.20\%  \\
        \midrule
                 & BBSR & 6.67\% & 93.51\% & 9.38\%  & 7.23\% & 98.43\% & 7.50\%  & 1.07\% & 87.92\% & 1.74\%  \\
        FLAME    & ABSR & 3.47\% & 89.59\% & 3.91\%  & 4.02\% & 90.12\% & 3.40\%  & 0.53\% & 81.16\% & 0.92\%  \\
                 & ACC  & 84.12\% & 74.35\% & 84.49\%  & 78.65\% & 74.73\% & 77.93\%  & 50.68\% & 40.80\% & 51.13\%  \\
        \midrule
                 & BBSR & 94.80\% & 14.34\% & 100.00\%  & 7.12\% & 6.27\% & 100.00\%  & 93.81\% & 4.01\% & 100.00\%  \\
        FLIP     & ABSR & 10.33\% & 7.25\% & 100.00\%  & 2.82\% & 3.48\% & 100.00\%  & 5.27\% & 1.80\% & 100.00\%  \\
                 & ACC  & 86.72\% & 74.93\% & \textbf{87.89\%}  & 81.02\% & 75.64\% & 82.24\%  & 59.22\% & 43.27\% & 59.14\%  \\  
        \midrule
                 & BBSR & \textbf{\phantom{0}4.59\%} & \textbf{\phantom{0}4.59\%} & \textbf{\phantom{0}4.83\%}  & \textbf{\phantom{0}2.51\%} & \textbf{\phantom{0}5.03\%} & \textbf{\phantom{0}3.23\%}  & \textbf{\phantom{0}0.95\%} & \textbf{\phantom{0}0.97\%} & 2.60\%  \\
        FedBAP (Ours)   & ABSR & \textbf{\phantom{0}2.16\%} & \textbf{\phantom{0}2.75\%} & \textbf{\phantom{0}2.21\%}  & \textbf{\phantom{0}1.84\%} & \textbf{\phantom{0}3.05\%} & \textbf{\phantom{0}2.26\%}  & 0.55\% & \textbf{\phantom{0}0.55\%} & 1.79\%  \\
                 & ACC  & \textbf{90.55\%} & \textbf{80.81\%} & 86.58\%  & \textbf{87.73\%} & \textbf{81.07\%} & \textbf{88.24\%}  & \textbf{63.61\%} & \textbf{46.51\%} & \textbf{64.08\%}  \\
        \bottomrule
    \end{tabular}
\end{table*}

\begin{table*}
    \centering
    \caption{Performance of different defenses on Fashion-MNIST.}
    \label{tab:fashion_mnist_defense}
    \begin{tabular}{l | S[table-format=3.2, table-space-text-post=\%] S[table-format=3.2, table-space-text-post=\%] S[table-format=3.2, table-space-text-post=\%] | S[table-format=3.2, table-space-text-post=\%] S[table-format=3.2, table-space-text-post=\%] S[table-format=3.2, table-space-text-post=\%] | S[table-format=3.2, table-space-text-post=\%] S[table-format=3.2, table-space-text-post=\%] S[table-format=3.2, table-space-text-post=\%]}
        \toprule
        & \multicolumn{3}{c|}{BadNets} & \multicolumn{3}{c|}{A3FL} & \multicolumn{3}{c}{CerP} \\
        \cmidrule(lr){2-4} \cmidrule(lr){5-7} \cmidrule(lr){8-10}
        & {BBSR} & {ABSR} & {ACC} & {BBSR} & {ABSR} & {ACC} & {BBSR} & {ABSR} & {ACC} \\
        \midrule
        FedAvg            & 99.98\% & 99.93\% & \textbf{\phantom{0}92.46\%} & 99.99\% & 99.97\% & 91.90\% & 100.00\% & 100.00\% & \textbf{\phantom{0}92.38\%} \\
        Krum              & 99.63\% & 22.47\% & 79.59\% & 99.76\% & 39.65\% & 77.51\% & 74.33\%  & 7.88\%   & 75.75\% \\
        MultiKrum         & 100.00\%& 99.73\% & 91.14\% & 99.97\% & 99.90\% & 91.57\% & 100.00\% & 99.99\%  & 91.82\% \\
        FLTrust           & 17.38\% & 13.41\% & 89.66\% & 98.53\% & 81.43\% & 89.93\% & 96.45\%  & 79.85\%  & 89.83\% \\
        RLR               & 100.00\%& 86.66\% & 90.94\% & 99.99\% & 86.66\% & 91.05\% & 100.00\% & 80.00\%  & 91.24\% \\
        FLAME             & 99.48\% & 95.99\% & 91.19\% & 99.98\% & 99.95\% & 91.12\% & 100.00\% & 100.00\% & 91.35\% \\
        FLIP              & 11.16\% & 1.19\%  & 91.92\% & 94.42\% & 15.36\% & \textbf{\phantom{0}92.02\%} & 100.00\% & 100.00\% & 92.21\% \\
        RoseAgg           & 99.68\% & 73.34\% & 91.18\% & 100.00\%& 99.71\% & 91.12\% & 100.00\% & 99.78\%  & 92.04\% \\
        Snowball          & 99.97\% & 99.88\% & 91.84\% & 100.00\%& 99.98\% & 91.57\% & 100.00\% & 99.98\%  & 91.67\% \\
        BackdoorIndicator & 99.93\% & 50.42\% & 88.32\% & 99.98\% & 99.93\% & 88.88\% & 100.00\% & 99.41\%  & 87.86\% \\
        FedBAP (Ours)     & \textbf{\phantom{0}1.77\%}  & \textbf{\phantom{0}0.42\%}  & 90.79\% & \textbf{\phantom{0}0.65\%}  & \textbf{\phantom{0}0.21\%}  & 90.83\% & \textbf{\phantom{0}0.43\%}   & \textbf{\phantom{0}0.14\%}   & 90.15\% \\
        \bottomrule
    \end{tabular}
\end{table*}

\begin{figure*}[t]
    \centering
    \begin{subfigure}[b]{0.24\linewidth}
        \centering
        \includegraphics[width=\linewidth]{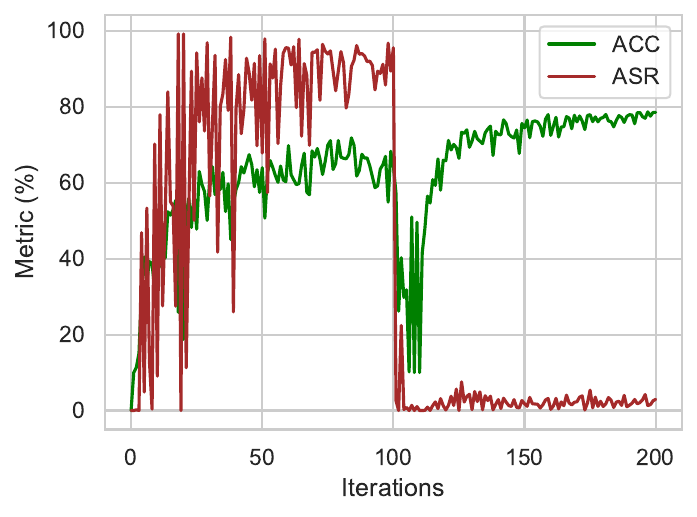}
        \caption{FedBAP (Ours)}
    \end{subfigure}
    \hfill
    \begin{subfigure}[b]{0.24\linewidth}
        \centering
        \includegraphics[width=\linewidth]{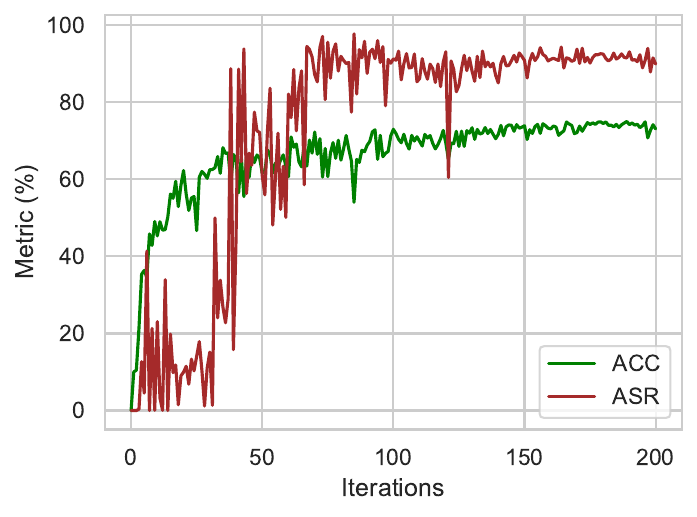}
        \caption{FedAvg}
    \end{subfigure}
    \hfill
    \begin{subfigure}[b]{0.24\linewidth}
        \centering
        \includegraphics[width=\linewidth]{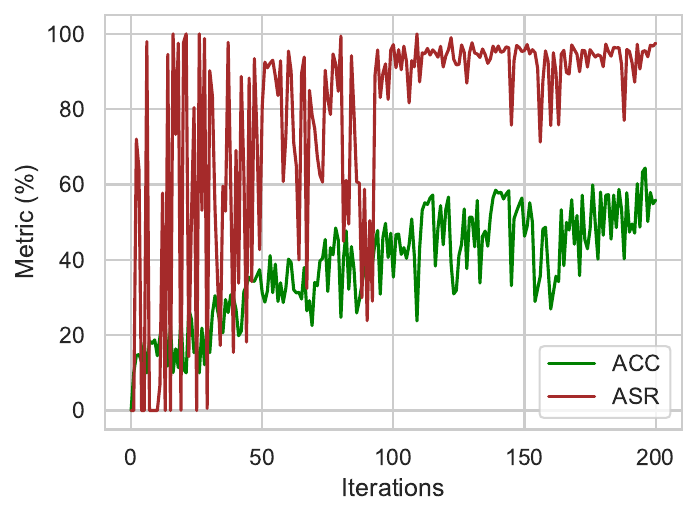}
        \caption{Krum}
    \end{subfigure}
    \hfill
    \begin{subfigure}[b]{0.24\linewidth}
        \centering
        \includegraphics[width=\linewidth]{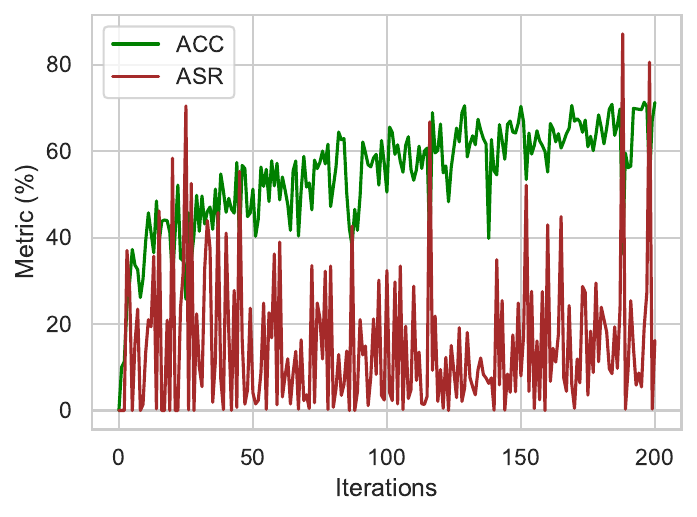}
        \caption{MultiKrum}
    \end{subfigure}

    \par\medskip

    \begin{subfigure}[b]{0.24\linewidth}
        \centering
        \includegraphics[width=\linewidth]{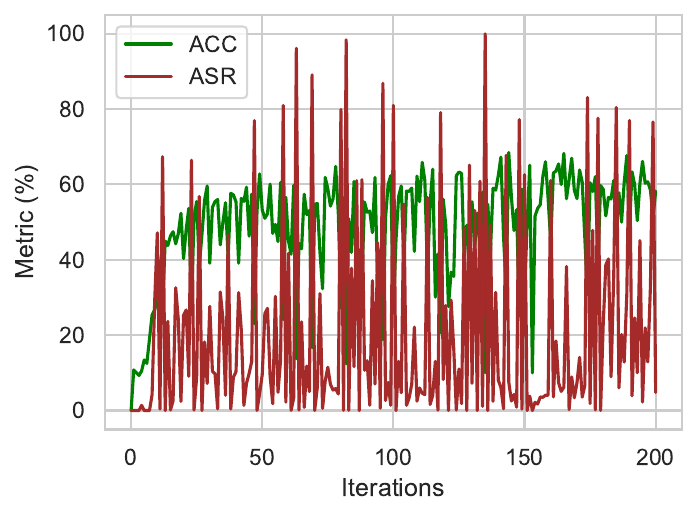}
        \caption{FLTrust}
    \end{subfigure}
    \hfill
    \begin{subfigure}[b]{0.24\linewidth}
        \centering
        \includegraphics[width=\linewidth]{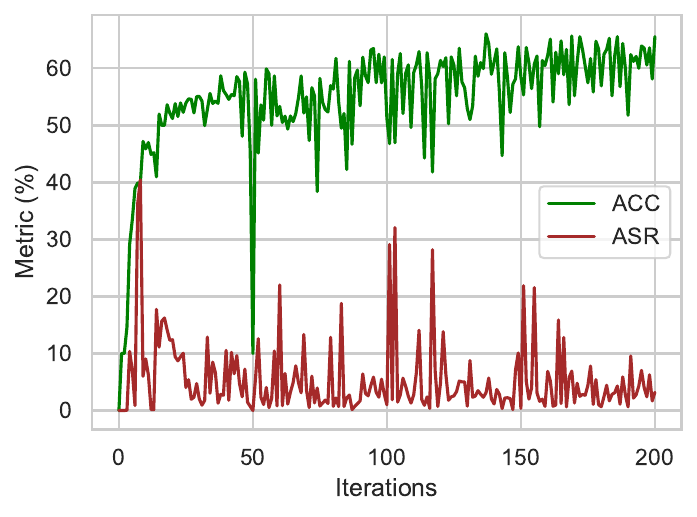}
        \caption{RLR}
    \end{subfigure}
    \hfill
    \begin{subfigure}[b]{0.24\linewidth}
        \centering
        \includegraphics[width=\linewidth]{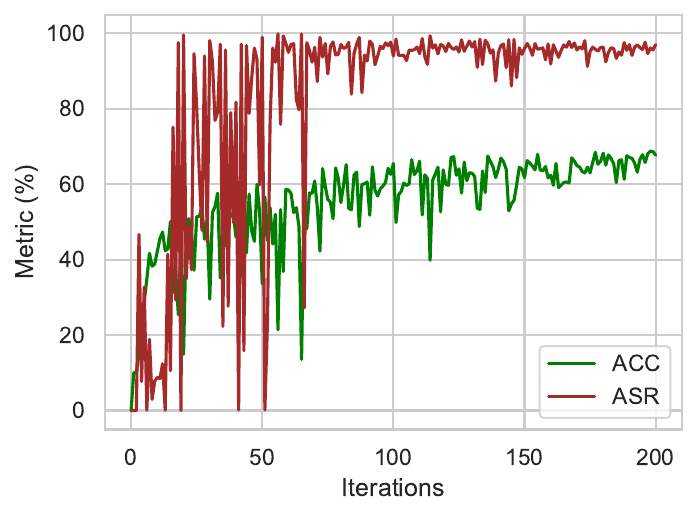}
        \caption{FLAME}
    \end{subfigure}
    \hfill
    \begin{subfigure}[b]{0.24\linewidth}
        \centering
        \includegraphics[width=\linewidth]{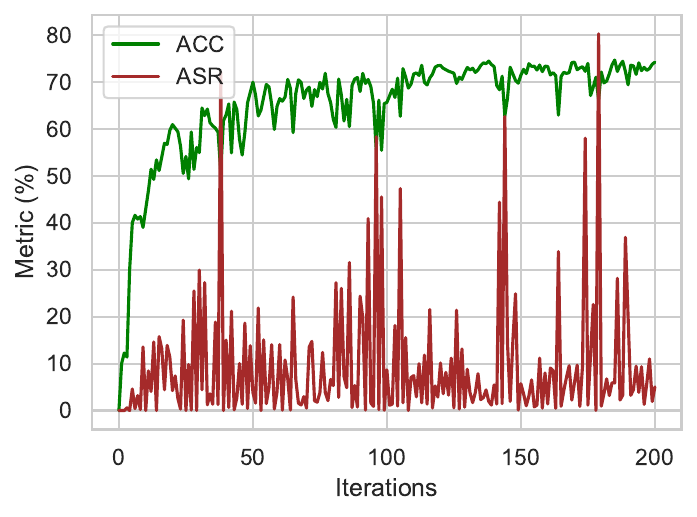}
        \caption{FLIP}
    \end{subfigure}
    
    \caption{ResNet-18 trained with different defenses on the non-IID CIFAR10 dataset against the LP attack.}
    \Description{}
    \label{lp}
\end{figure*}

\begin{figure*}[t]
    \centering
    \begin{subfigure}[b]{0.24\linewidth}
        \centering
        \includegraphics[width=\linewidth]{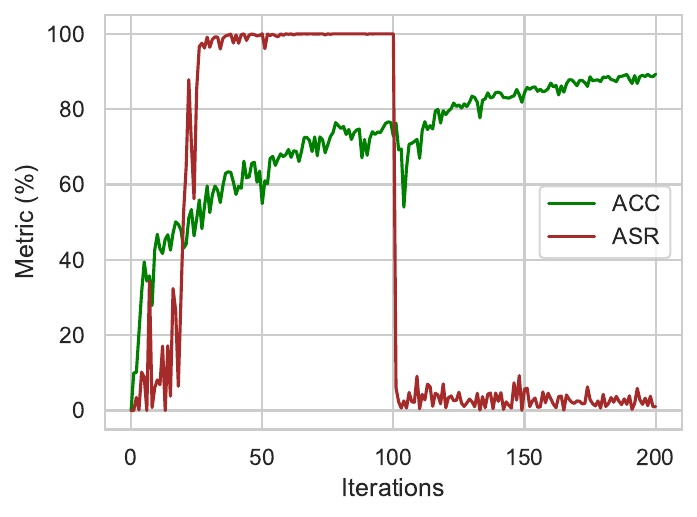}
        \caption{FedBAP (Ours)}
    \end{subfigure}
    \hfill
    \begin{subfigure}[b]{0.24\linewidth}
        \centering
        \includegraphics[width=\linewidth]{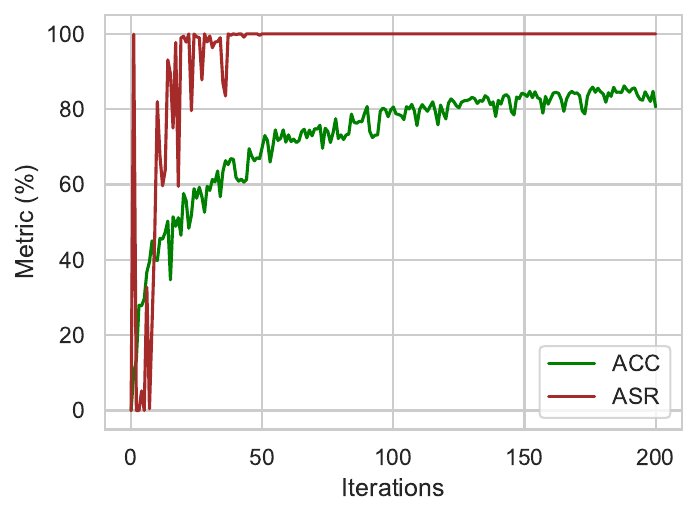}
        \caption{FedAvg}
    \end{subfigure}
    \hfill
    \begin{subfigure}[b]{0.24\linewidth}
        \centering
        \includegraphics[width=\linewidth]{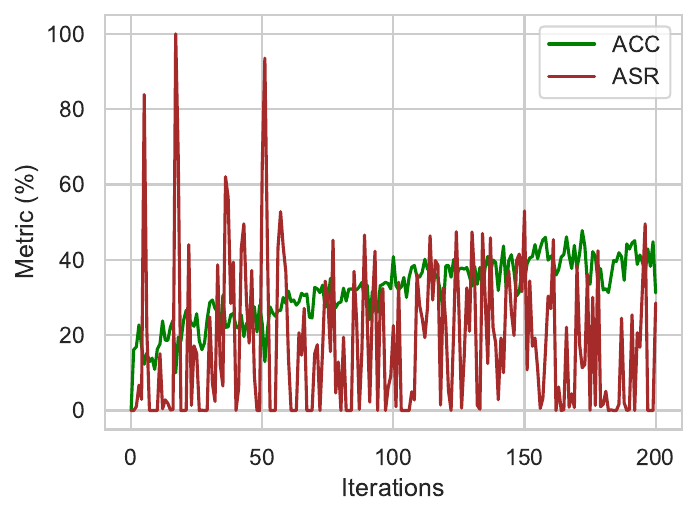}
        \caption{Krum}
    \end{subfigure}
    \hfill
    \begin{subfigure}[b]{0.24\linewidth}
        \centering
        \includegraphics[width=\linewidth]{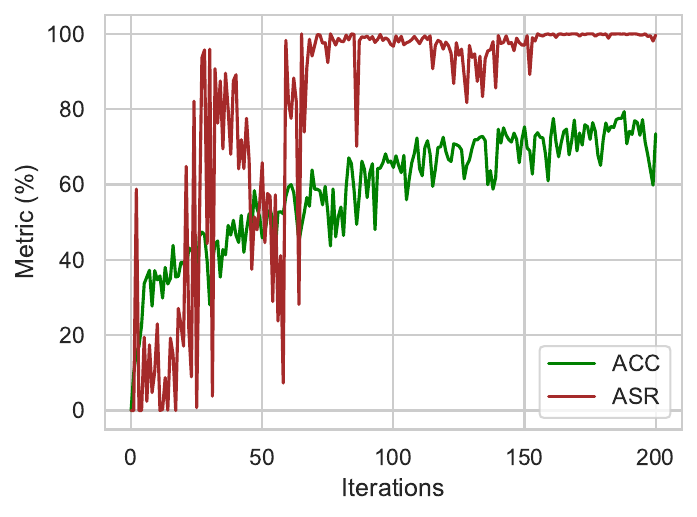}
        \caption{MultiKrum}
    \end{subfigure}

    \par\medskip

    \begin{subfigure}[b]{0.24\linewidth}
        \centering
        \includegraphics[width=\linewidth]{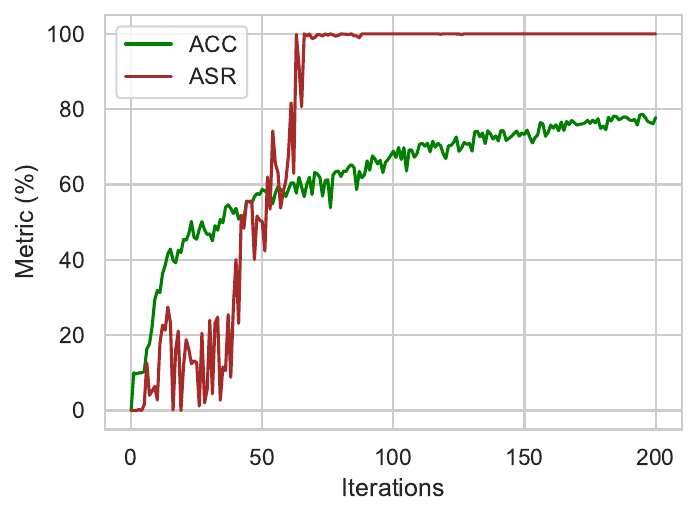}
        \caption{FLTrust}
    \end{subfigure}
    \hfill
    \begin{subfigure}[b]{0.24\linewidth}
        \centering
        \includegraphics[width=\linewidth]{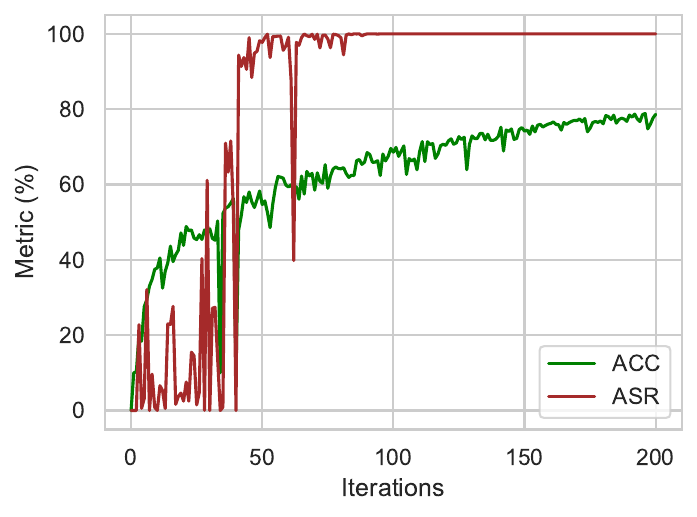}
        \caption{RLR}
    \end{subfigure}
    \hfill
    \begin{subfigure}[b]{0.24\linewidth}
        \centering
        \includegraphics[width=\linewidth]{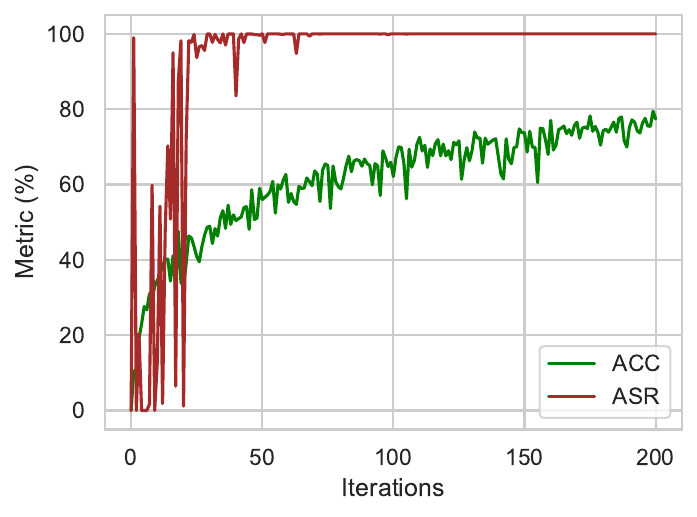}
        \caption{FLAME}
    \end{subfigure}
    \hfill
    \begin{subfigure}[b]{0.24\linewidth}
        \centering
        \includegraphics[width=\linewidth]{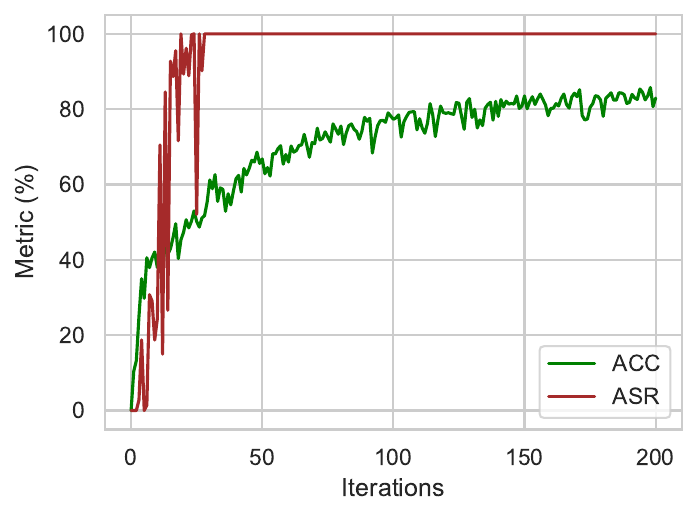}
        \caption{FLIP}
    \end{subfigure}
    \caption{ResNet-18 trained with different defenses on the non-IID CIFAR10 dataset against the A3FL attack.}
    \Description{}
    \label{a3fl}
\end{figure*}

\section{Baseline Defenses}

\begin{itemize}
    \item \textbf{FedAvg} \cite{mcmahan2017communication} is one of the most popular algorithms in FL. It is designed to address the challenges of training machine learning models in a decentralized manner, where data is distributed across multiple devices and cannot be shared due to privacy concerns. The core idea of FedAvg is to perform local model updates on each participating device and then aggregate these updates to form the global model. Specifically, each client computes model updates by training locally on its own data, and after a few local updates, the models are averaged on the central server. This process reduces the need for sharing raw data, as only model updates are exchanged.

    \item \textbf{Krum} \cite{blanchard2017machine} is a Byzantine-resilient aggregation algorithm designed to safeguard distributed learning, particularly in the context of Stochastic Gradient Descent. In a distributed system with $n$ workers, Krum can tolerate up to $f$ Byzantine workers. The algorithm works by selecting the model update that is closest to the majority of other updates, effectively identifying and excluding those that deviate significantly, which are assumed to be malicious. This approach ensures that even with the presence of faulty or adversarial updates, the global model can still converge correctly.

    \item \textbf{MultiKrum} is an extension of the Krum algorithm designed to provide greater robustness in distributed learning systems, particularly against Byzantine faults. While Krum aggregates model updates by selecting the one closest to the majority, MultiKrum improves upon this by considering multiple candidates rather than just one. Specifically, MultiKrum selects the $k$ closest updates, where $k$ is a predetermined number based on the level of fault tolerance required. By doing so, it ensures that even if more than one Byzantine worker is present, the global model can still be updated reliably. This multi-candidate approach increases resilience, offering better protection against adversarial behavior compared to the single-update selection in Krum.

    \item \textbf{FLTrust} \cite{cao2020fltrust} is a Byzantine-robust FL method that bootstraps trust from the server side by leveraging a small, clean root dataset. Unlike other defenses that solely rely on statistical filtering of client updates, FLTrust maintains a server model trained on trusted data and uses it to evaluate the alignment of client updates. It assigns trust scores based on the directional similarity between client updates and the server model update, normalizes all updates to limit their impact, and then performs trust-weighted aggregation. This approach significantly improves robustness against strong adaptive attacks, even when a large fraction of clients are malicious.

    \item \textbf{RLR} \cite{ozdayi2021defending} is a defense mechanism designed to mitigate backdoor attacks in FL. Instead of filtering or removing suspicious client updates, RLR takes a different approach by dynamically adjusting the server-side learning rate for each dimension of the model update. This adjustment is based on the degree of agreement among clients on the sign of the update in each dimension. The core intuition is that dimensions with high disagreement are more likely to be manipulated by malicious clients, and thus should be updated more conservatively. By selectively shrinking the learning rate in these contentious dimensions, RLR effectively suppresses the influence of poisoned updates while still allowing benign information to be aggregated. RLR does not require knowledge of which clients are malicious, and can be seamlessly integrated into standard FL protocols with minimal overhead. 

    \item \textbf{FLAME} \cite{nguyen2022flame} is a defense framework against backdoor attacks in FL that aims to eliminate malicious behavior while preserving model utility. Unlike filtering-based methods or differential privacy approaches that may suffer from limited threat coverage or significant accuracy degradation, FLAME estimates the minimal amount of noise required to remove backdoors. It employs model clustering and weight clipping to reduce the necessary noise injection, thereby maintaining benign performance.

    \item \textbf{FLIP} \cite{zhang2022flip} is a novel defense strategy based on trigger reverse engineering to mitigate backdoor attacks in FL. Unlike previous robust aggregation or certified robustness methods, FLIP focuses on hardening benign clients and analyzes the theoretical relationship between cross-entropy loss, attack success rate, and clean accuracy. It guarantees reduced attack success without harming benign performance. 

\end{itemize}

\section{Additional Experimental Result}

Table \ref{tab:summary_results_iid} presents the performance of different defenses against three backdoor attacks across various model architectures and datasets in IID settings. The results in Table \ref{tab:summary_results_iid} clearly illustrate the performance of different defense strategies against backdoor attacks on IID datasets. The result demonstrate that FedBAP consistently shows superior performance in both resisting backdoor attacks and maintaining accuracy across different settings.

Figure \ref{lp}, \ref{a3fl} illustrates the BSR and ACC curves of different defenses against the LP and A3FL attacks on ResNet-18 with CIFAR-10. The curves in Figures \ref{lp} and \ref{a3fl} demonstrate that FedBAP effectively keeps the BSR within a low range while maintaining stability, without any sudden spikes. This indicates that FedBAP can consistently control the impact of attacks and avoid instability in the defense performance. Such stability is crucial for ensuring the reliability and long-term effectiveness of the defense system.

To further validate the effectiveness and generalizability of FedBAP, we extend our experiments to the Fashion-MNIST dataset. We additionally incorporate the CerP ~\cite{lyu2023poisoning} attack and compare FedBAP with recently proposed defenses, including RoseAgg ~\cite{yang2024roseagg}, Snowball ~\cite{qin2024resisting}, and BackdoorIndicator ~\cite{li2024backdoorindicator}. Specifically, we adopt ResNet-18 as the global model architecture, set the malicious client proportion to 30\%, the total number of global communication rounds to 100, and the defense start round in FedBAP to 85. As shown in Table \ref{tab:fashion_mnist_defense}, FedBAP demonstrates strong and consistent performance across all attack types.

\end{document}